\documentclass[preprint]{aastex61}
\usepackage{amsmath}

\shorttitle{Flare Prediction Using Photospheric and Coronal Image Data}
\shortauthors{Jonas et al.}
\usepackage{natbib}
\usepackage{hyperref}
\usepackage{booktabs}
\bibpunct{(}{)}{;}{}{,}{,}
\usepackage{tcolorbox}
\newcommand{\Reals}{\mathbb{R}}
\newcommand{\vect}[1]{\boldsymbol{#1}}
\newcommand{\sdoseries}[1]{\texttt{#1}}
\newcommand{\segment}[1]{\texttt{#1}}

\let\oldAA\AA
\renewcommand{\AA}{\text{\normalfont\oldAA}}

\begin{document}
\DeclareGraphicsRule{.ai}{pdf}{.ai}{}
\title{Flare Prediction Using Photospheric and Coronal Image Data}
\author[0000-0001-6159-3450]{Eric Jonas}
\affiliation{Department of Electrical Engineering and Computer Science, University of California, Berkeley, CA}
\author[0000-0002-5662-9604]{Monica Bobra}
\affiliation{W. W. Hansen Experimental Physics Laboratory, Stanford University, Stanford, CA}
\author{Vaishaal Shankar}
\affiliation{Department of Electrical Engineering and Computer Science, University of California, Berkeley, CA}
\author[0000-0001-9130-7312]{J. Todd Hoeksema}
\affiliation{W. W. Hansen Experimental Physics Laboratory, Stanford University, Stanford, CA}
\author{Benjamin Recht}
\affiliation{Department of Electrical Engineering and Computer Science, University of California, Berkeley, CA}

\begin{abstract}

\noindent The precise physical process that triggers solar flares is not currently understood. Here we attempt to capture the signature of this mechanism in solar image data of various wavelengths and use these signatures to predict flaring activity. We do this by developing an algorithm that [1] automatically generates features in 5.5 TB of image data taken by the Solar Dynamics Observatory of the solar photosphere, chromosphere, transition region, and corona during the time period between May 2010 and May 2014, [2] combines these features with other features based on flaring history and a physical understanding of putative flaring processes, and [3] classifies these features to predict whether a solar active region will flare within a time period of $T$ hours, where $T$ = 2 and 24. This type of machine-learning algorithm is conceptually similar to a single-layer Convolutional Neural Network (CNN) with pre-specified filters that is trained using a linear classifier. Such an approach may be useful since, at the present time, there are no physical models of flares available for real-time prediction. We find that when optimizing for the True Skill Score (TSS), photospheric vector magnetic field data combined with flaring history yields the best performance, and when optimizing for the area under the precision-recall curve, all the data are helpful. Our model performance yields a TSS of $0.84 \pm 0.03$ and $0.81 \pm 0.03$ in  the $T$ = 2 and 24 hour cases, respectively,  and a value of $0.13 \pm 0.07$ and $0.43 \pm 0.08$ for the area under the precision-recall curve in the $T$ = 2 and 24 hour cases, respectively. These relatively high scores are similar to, but not greater than, other attempts to predict solar flares. Given the similar values of algorithm performance across various types of models reported in the literature, we conclude that we can expect a certain baseline predictive capacity using these data. This is the first attempt to predict solar flares using photospheric vector magnetic field data as well as multiple wavelengths of image data from the chromosphere, transition region, and corona.

\pagebreak
\end{abstract}

% commenting out fixme
%\listoffixmes
%\tableofcontents

\section{Introduction}
\label{section:intro}

The rapid reconfiguration of the solar magnetic field during a flare triggers physical processes that can, over the time period of seconds to minutes, emit wavelengths spanning 17 orders of magnitude, from radio waves to gamma-rays \citep{schwenn06}. This radiation can, in turn, affect the Earth. As such, it is critical to study the solar magnetic field in order to understand, and ultimately predict, flares. 

Since we first observe emerging magnetic flux on the photosphere, and since, until recently, it was only possible to map the solar magnetic field at the photosphere, many flare prediction models use photospheric magnetic field data for their predictive task (e.g. \citealt{lb03}, \citealt{welsch09}). This is a logical starting point, since energy budgets of active regions can be adequately measured using photospheric data \citep{ForbesPriest2002} and many features of the photospheric magnetic field, such as the presence of polarity inversion lines, are strongly correlated with flaring activity (e.g. \citealt{zirin93}, \citealt{jing06}, \citealt{schrijver07}, \citealt{mason10}, \citealt{falconer12}). However, observations of solar flares show dynamic behavior in the coronal magnetic field, particularly in the transition region (e.g. \citealt{benz17}), while changes in the photospheric magnetic field are near the detection limit \citep{sudol05}. As such, a more complete approach is to build a predictive model that uses multiple wavelengths of solar image data at many heights.

Many studies show a statistical correlation between flare productivity and features in multiple wavelengths of solar image data. For example, \citet{canfield99} demonstrated that a large active region, as observed on the photosphere, that is co-temporal with a sigmoidal feature in the X-ray solar corona is likely to lead to an eruption. \citet{su07} showed that over the course of a two-ribbon flare, or a flare with two ribbon-like signatures in the Transition Region and Coronal Explorer (TRACE) Extreme Ultraviolet (EUV) or Ultraviolet (UV) passbands, both ribbons move outward from the underlying polarity inversion line on the photosphere or chromosphere 86\% of the time. These examples are two of many, and a complete review can be found in \citet{benz17} and \citet{fletcher11}. Few studies, however, attempt to predict flares using multiple wavelengths of solar image data. To date, \citet{nishizuka17} is only study that has used both photospheric vector magnetic field data and chromospheric data to predict solar flares. This study used various parameterizations of the photospheric vector magnetic field and chromospheric data such that each active region can be described by several numbers. 

While it is common to parameterize magnetic field data to predict flares, there exists considerable debate on the best way to go about it. For example, some argue that the magnetic field near the polarity inversion line is most relevant (e.g. \citealt{schrijver07}), while others argue that the interconnectedness of the field may be more relevant (e.g. \citealt{george07}). Either way, choices must be made to engineer the features that go into any predictive model. 

There are also many predictive models to choose from. A machine learning algorithm, which is favorable in cases where the amount of data is large, is one way to predict flares. The simplest approach is to use a subset of such algorithms known as binary classifiers, which can predict whether or not an active region will flare within a certain time interval. To do this, the algorithm develops a framework by characterizing some fraction of the data for which it already knows the answer. For example, consider a data set consisting of two features (e.g. the size of an active region and the total flux contained within this active region) and two labels (active regions that produced flares and those that did not). In this case, the algorithm learns which combination of features is most likely correlated with flaring activity. In other words, the algorithm assigns weights to each feature. These weights form the basis for predicting the outcome of a future example, which contains two features but no labels. 

Machine learning algorithms have been used in many solar flare prediction studies (e.g. \citealt{song09}, \citealt{yu09}, \citealt{yuan10}, \citealt{ahmed13}, \citealt{boucheron15}, \citealt{nishizuka17}), but the features that drive these models have traditionally been hand-engineered based on physical insights. Another approach, which is popular in the computer vision community, involves extracting relatively simple and generic features from the image data, and allowing the learning algorithm to pick the most useful ones. Deriving these features usually involves convolving, thresholding, and downsampling image data using various filters. In this case, the algorithm learns which combination of filters will distinguish images of flaring active regions from images of non-flaring ones. One practical advantage of such an algorithm is that it allows the user to easily incorporate various types of image data with limited pre-processing. 

Such an approach may be useful since, at the present time, there are no physical models of flares available for real-time prediction. To this end, we develop a machine-learning algorithm and use it with solar image data from the Solar Dynamics Observatory to predict solar flares. This is the first attempt to predict solar flares using photospheric vector magnetic field data as well as multiple wavelengths of image data from the chromosphere, transition region, and corona. 

This paper is organized as follows: Section \ref{section:data} describes the data, Section \ref{section:task} describes the prediction task, Section \ref{section:algorithm} describes in the predictive model, Section \ref{section:results} describes the results, and we discuss in Section \ref{section:conclusion}. 

%All of the data and codes used in this study are publicly available at [URL]\footnote{To the referee: We will make our code publicly available when the paper is accepted for publication. For now, you can access the code here: .}.

\section{Data}
\label{section:data}

\begin{figure*}
\renewcommand{\tabcolsep}{0.0018\textwidth}
\begin{tabular}{cc}
\includegraphics[angle=0,width=0.496\textwidth]{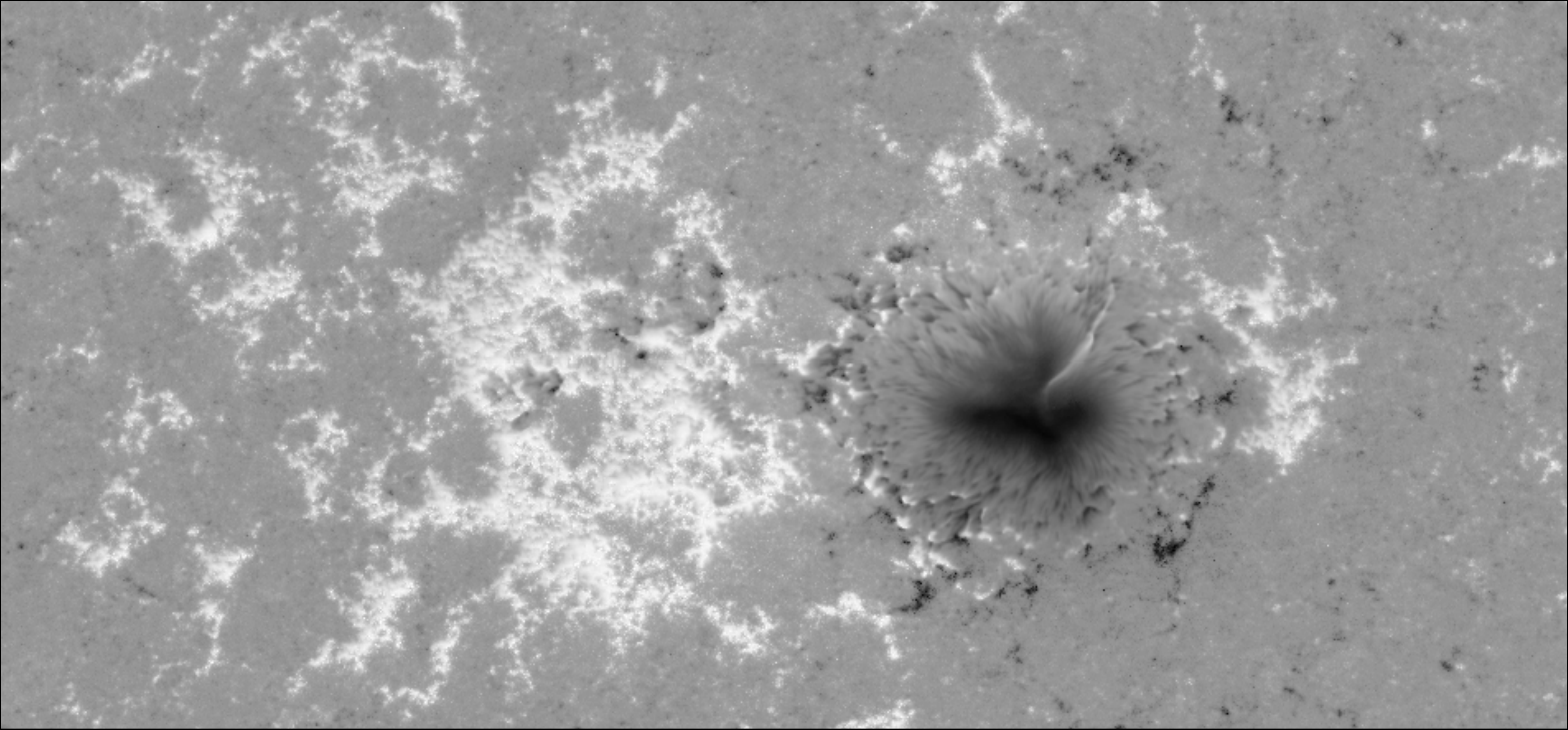} &
\includegraphics[angle=0,width=0.496\textwidth]{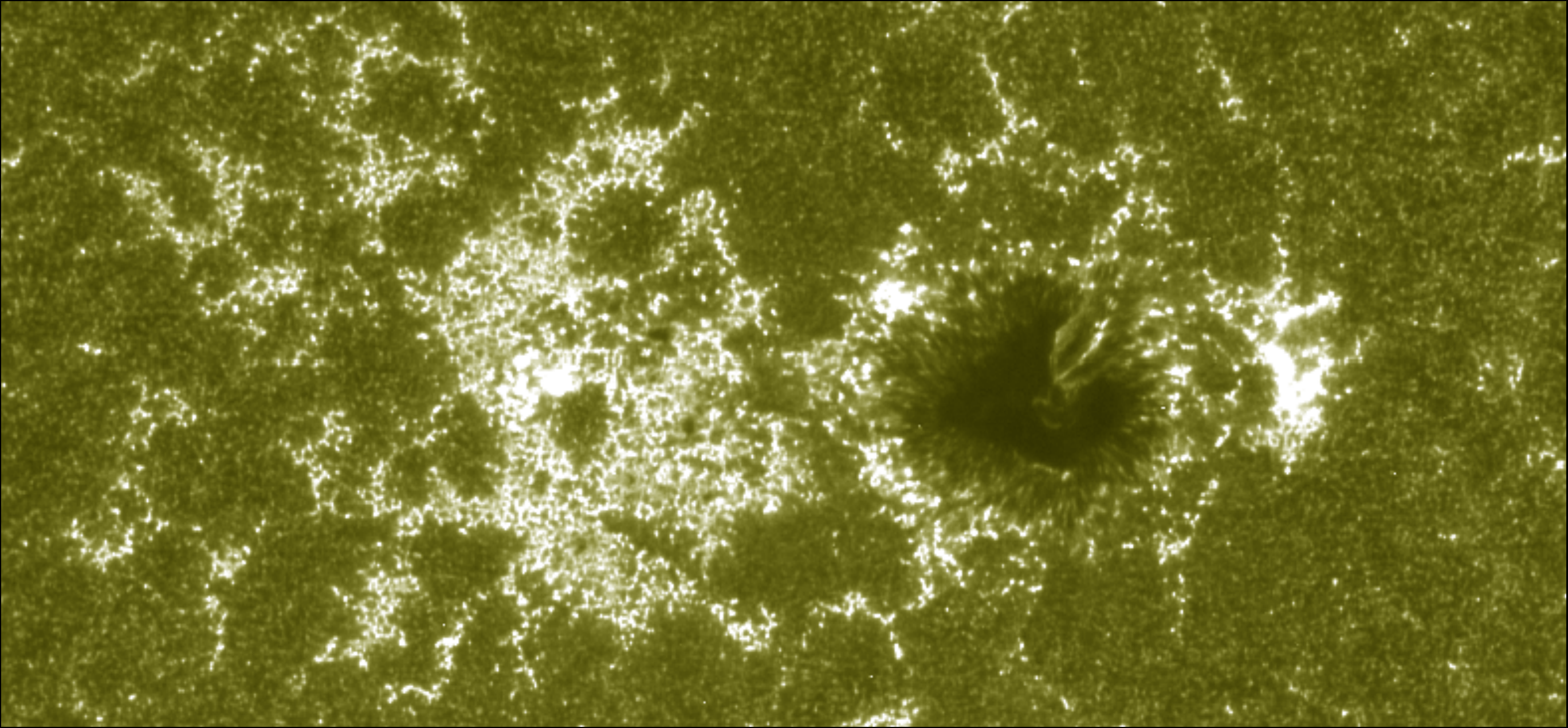} \\
\includegraphics[angle=0,width=0.496\textwidth]{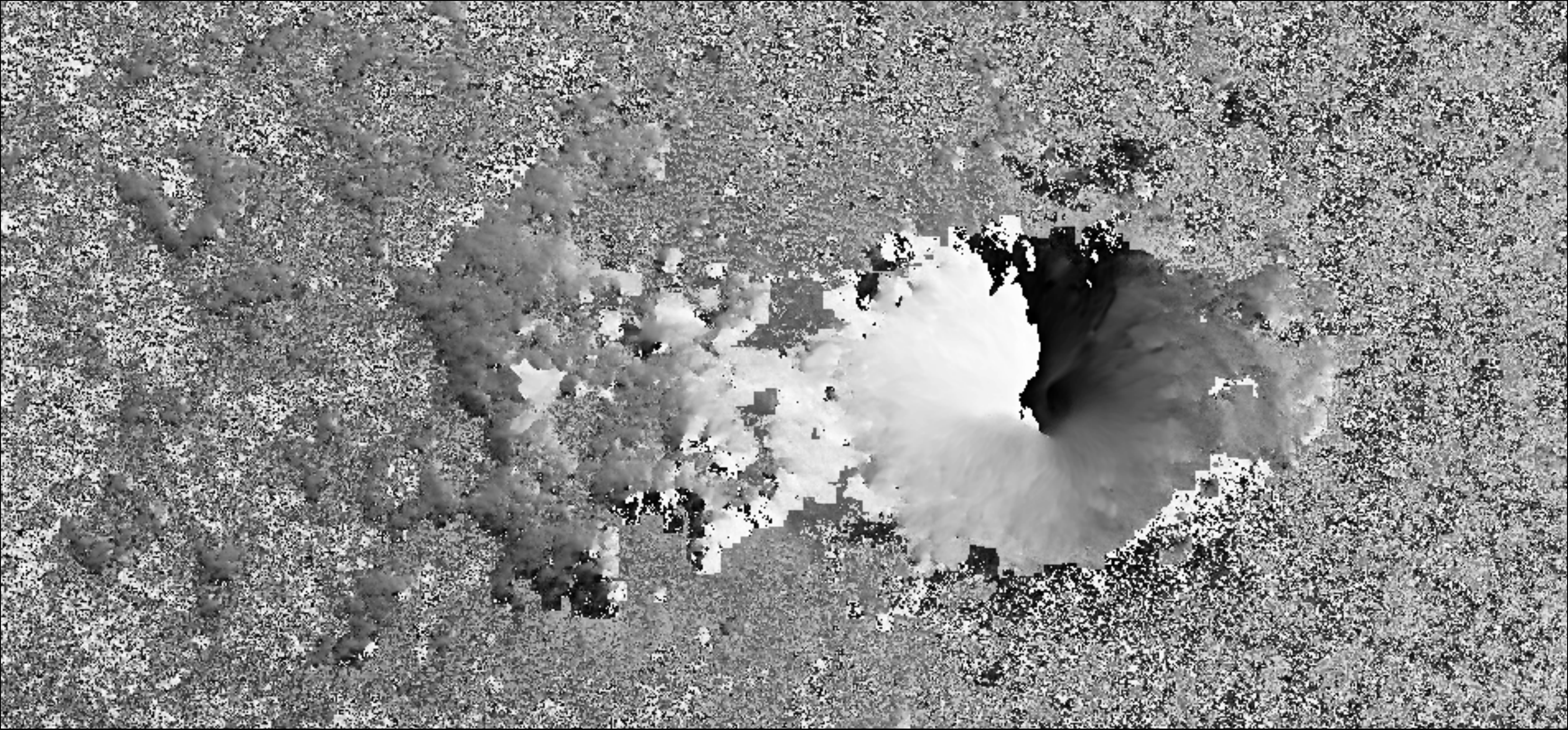} &
\includegraphics[angle=0,width=0.496\textwidth]{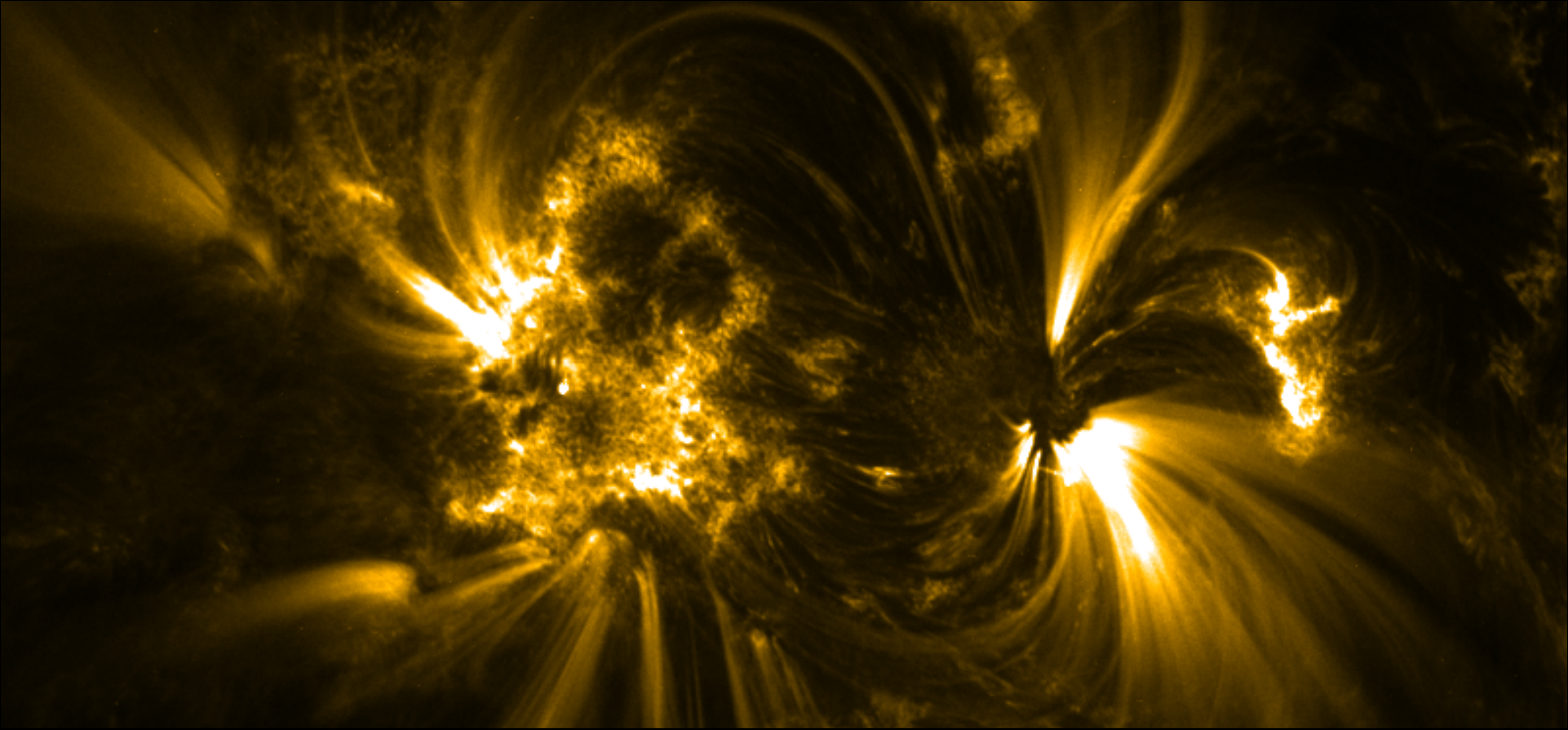} \\
\includegraphics[angle=0,width=0.496\textwidth]{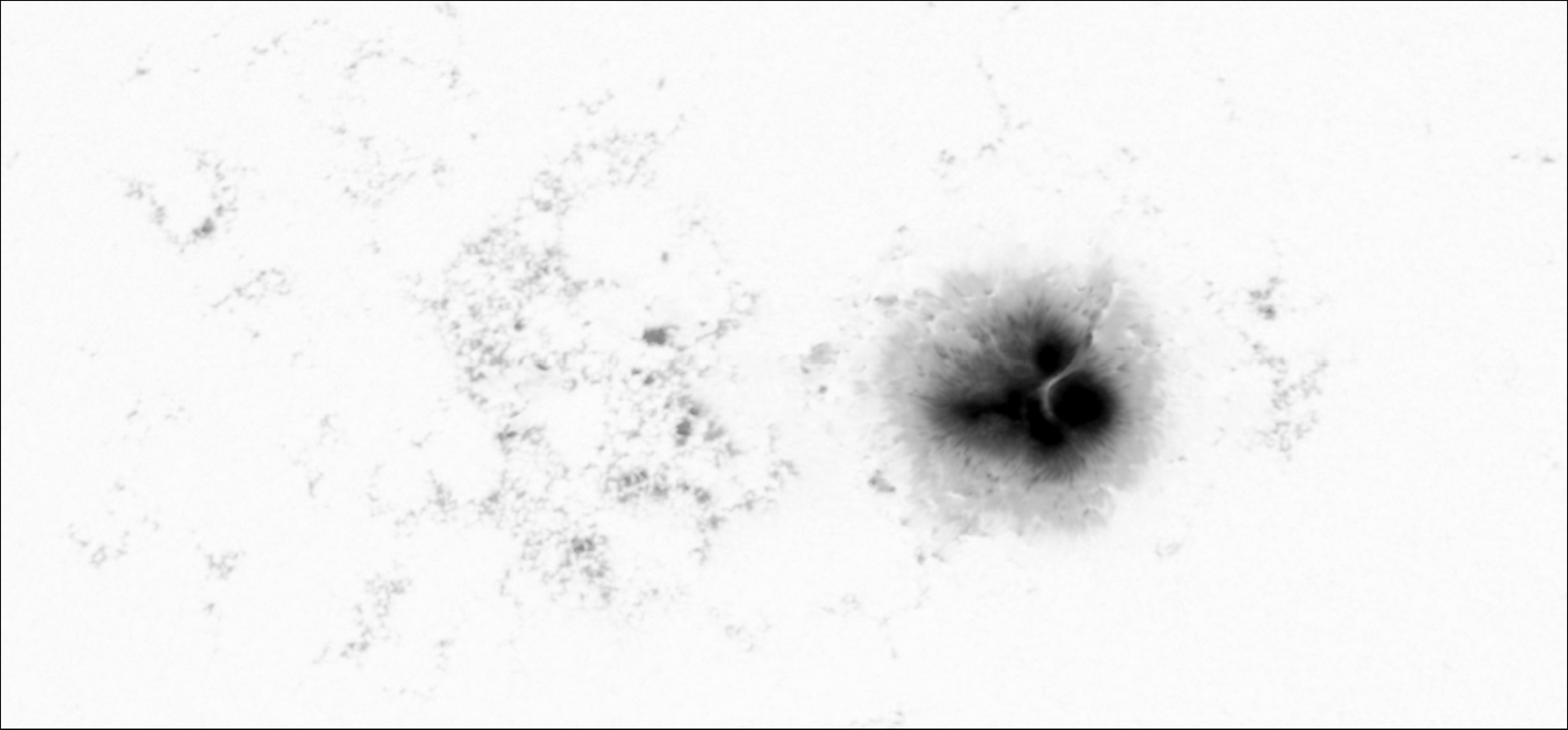} &
\includegraphics[angle=0,width=0.496\textwidth]{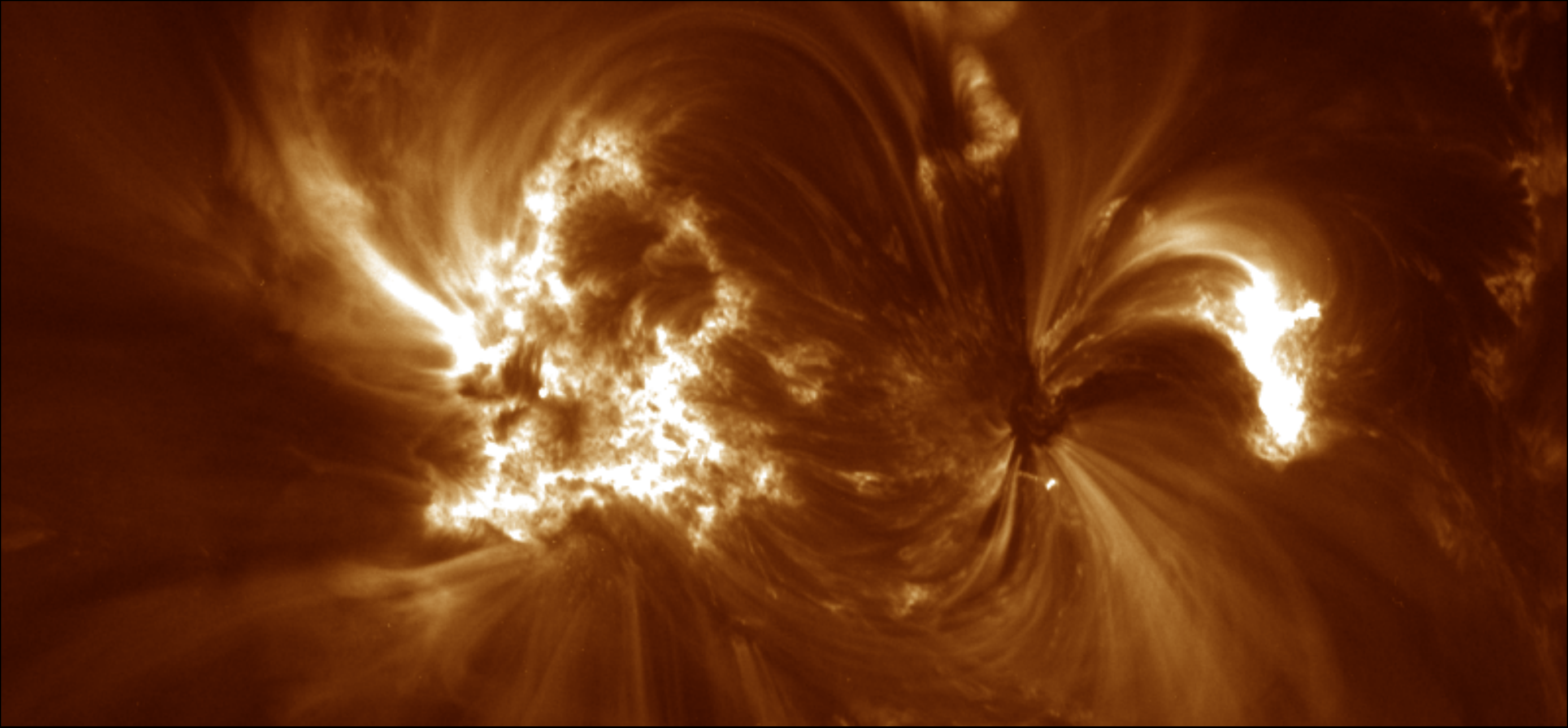} \\
\end{tabular}
\caption{The first column shows each of the components of the vector magnetic field data -- the inclination angle, azimuthal angle, and total field strength -- for NOAA Active Region 12529 on April 14, 2016 at 00:00 TAI. This active region produced a M6.7-class flare two days later, on April 18, 2016 at 00:29 TAI. The second column shows images of the chromosphere, in 1600 \AA, the transition region, in 171 \AA, and the corona, in 193 \AA, for the same active region. These three images are not exactly co-temporal with the magnetic field data; they are taken 5, 11, and 7 seconds afterward, respectively. The color table for the magnetic field data is scaled from $0^\circ$ to $180^\circ$ for the inclination angle, $0^\circ$ to $360^\circ$ for the azimuthal angle, and between $\pm$ 2500 Gauss for the field strength. The color table for the chromospheric data is scaled from 0 to 400 DN; the remaining two EUV channels are scaled from 0 to 2500 DN and 0 to 3500 DN. All of the images are shown in CCD coordinates and were remapped to heliographic Cylindrical Equal-Area coordinates and decomposed into B$_{r}$, B$_{\theta}$, and B$_{\phi}$ using the equations outlined in \citet{sun13}. The center of this active region is at $344.1^\circ$ in Carrington longitude and $-1.6^\circ$ in latitude during Carrington Rotation 2176.}
\label{fig:example}
\end{figure*}

\subsection{Image Data}
\label{subsection:image_data}
The Solar Dynamics Observatory \citep{pesnell12}, which has been taking data continuously since May 2010, has two imaging instruments onboard: the Helioseismic and Magnetic Imager (HMI), which is the first space-based instrument to map the full-disk photospheric vector magnetic field \citep{schou12}, and the Atmospheric Imaging Assembly, which images the chromosphere, transition region, and corona in 9 different UV and EUV wavelengths \citep{lemen12}. These data are publicly available at the Joint Science Operations Center (see Table \ref{tab:datavolume}). HMI vector magnetic field data are taken continuously and averaged to a cadence of 12 minutes; AIA UV and EUV data are taken continuously at a cadence of 24 and 12 seconds, respectively. As such, any given HMI vector magnetic field image will be co-temporal with an AIA image within 24 seconds.

The HMI team also developed a higher-level data pipeline which automatically detects active regions in the full-disk image data. These automatically-detected active regions are referred to as HMI Active Region Patches, or HARPs, and available as bitmap arrays \citep{turmon10}.  Like NOAA active regions, multiple HARPs can appear on the solar disk at the same time. However, since HARPs are detected from line-of-sight magnetic field data, they capture more magnetic activity than the NOAA active region database. As such, there is no one-to-one correspondence between a NOAA active region number and HARP number (see Section 2 of \citealt{Bobra14} for more details). 

This pipeline then uses these HARP bitmaps as a template to extract corresponding vector magnetic field maps of each active region every 12 minutes throughout its lifetime \citep{Bobra14}. These maps, together with metadata that describe physical properties of the photospheric magnetic field, are called Space-weather HMI Active Region Patches (SHARP). A list of these physical quantities can be found in Table \ref{tab:params}. The three components of the photospheric vector magnetic field -- the azimuth, inclination, and strength -- are shown in the first column of Figure \ref{fig:example}. The ambiguity in the azimuthal component of the magnetic field is resolved using the minimum-energy solution \citep{metcalf94}, and the results of this solution are available as a bitmap array. We take the SHARP partial-disk vector magnetic field maps from the time period between May 2010 and May 2014, which contains 2253 active regions tracked throughout their lifetime, as our initial dataset. 

We then reject all records in this dataset where either of the following conditions are true: [1] The absolute value of the radial velocity of SDO is greater than 3500 m/s at the location of the HARP (see Section 7.1.2 of \citealt{hoeksema14} about the periodicity in magnetic field strength due to the orbital velocity of SDO), or [2] the HMI data are of low quality (see Section Appendix A of \citealt{hoeksema14} about HMI data quality). This process leaves 2182 active region time series where an average of 88\% of the time steps are retained. 

We use these partial-disk vector magnetic field maps to identify the corresponding areas in the full-disk AIA 1600 \AA, 171 \AA, and 193 \AA{} image data, thereby supplementing the output of this pipeline with partial-disk images of the chromosphere, transition region, and corona (see Figure \ref{fig:example}). We compensate for the degradation of the photon transmission through the AIA filters over the course of the SDO mission by normalizing the intensity observed in each channel to that which was observed in May 2010. The resulting data set, which spans the time period between May 2010 and May 2014, contains approximately 5 million images of 2182 active regions.

\subsection{Flare Data}
The X-Ray Sensor (XRS) instrument aboard the Geostationary Operational Environmental Satellite (GOES) measures the solar integrated X-Ray flux in two broadband channels: 1-8 \AA{} and 0.5 - 4 \AA. Various incarnations of XRS instruments on multiple GOES satellites have been making these measurements since 1974 (\citealt{garcia94}, \citealt{hanser96}). From these data, the GOES team generates a list of solar flares and attempts to assign each flare to a NOAA active region. Flares with GOES X-Ray flux values greater than $10^{-6}$ Watts/m$^2$ are classified as C-class; those with values above $10^{-5}$ and $10^{-4}$ Watts/m$^2$ are classified as M- and X-class, respectively. The NOAA Space Weather Prediction Center (SWPC) sends alerts when flares exceed the $5$x$10^{-5}$, or M5.0-class, level. Due to the constancy and longevity of the XRS measurements, the GOES event list has become the standard solar flare catalog. 

We query this flare catalog using SunPy \citep{sunpy} and select all flares greater than C1.0-class. From this initial list, we discard flares that are not associated with both a NOAA active region number and a HARP number. This leaves us with 3836 C-class flares, 406 M-class flares, and 30 X-class flares. It is worth noting that these numbers are relatively low, given the unusually quiet nature of solar cycle 24. 

\section{Prediction Task}
\label{section:task}

\begin{figure}
\centering
\includegraphics[angle=0,width=0.5\textwidth]{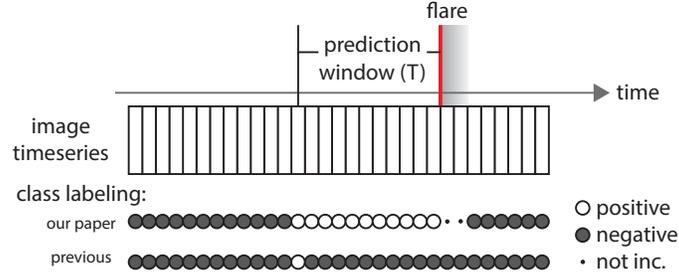} 
\caption{For each active region, we have a series of observations sampled at a 12 minute cadence. We are interested determining whether an active region will produce an M- or X-class flare within the next $T$ hours. This is a much more realistic problem than previous attempts to predict whether an active region will flare in exactly $T$ hours. We ascribe data taken $T$ hours prior to a flare to the positive class and all other data to the negative class.}
\label{fig:prediction_task}
\end{figure}

Our goal is to use past observations of an active region to predict its future flaring activity. We choose to model our problem as a binary classification task: Will this active region produce an M- or X-class flare within the next $T$ hours? For this study, we chose two values for $T$: 2 and 24. Predicting whether an active region will flare within 24 hours is common in the literature (e.g. \citealt{welsch09}, \citealt{ahmed13}), and a 2-hour prediction window may capture EUV emission during the preflare phase. Thus, we ascribe data taken $T$ hours prior to a flare to the positive class and all other data to the negative class. This task is illustrated in Figure \ref{fig:prediction_task}. Since flares are not instantaneous, we discard data during the flash phase of the flare.

Our definition of positive and negative class results in a substantial class imbalance. Since most active regions do not produce flares, far more observations are members of the negative class than the positive one. For the 2-hour prediction window, we had 1,252,111 observations in the negative class and 3,255 in the positive one, which is a ratio of 1:385. For the 24-hour prediction window, we had 1,232,862 observations in the negative class and 23,178 in the positive one, which is a ratio of 1:53. 

Ultimately, we are interested developing a real-time predictive model. In other words, we are interested in predicting \textit{future} flaring activity given current solar data. To get a sense for how our algorithm will perform on yet-unseen data, we simulate this process by splitting our data into two subsets. The first subset, or training set, contains the data our algorithm will learn from. The other subset, or testing set, is used to evaluate our algorithm. In the machine learning literature, this is known as cross-validation. Here we always use a test set size that constitutes approximately 20\% of the data. 

The performance of our algorithm may vary depending on precisely which data is included in the train and the test sets. To account for this, we generate up to 100 different 80\%-train and 20\%-test example experiments, or folds, of our data. Each fold is thus a different mixture of training and testing data, enabling a bootstrap-like estimate of our algorithm's performance. 

Note that correctly segregating this data into training and testing subsets can be difficult. In our experiments, we perform this segregation by active region -- that is, we place all the observations of a given active region in either the training or testing set. This ensures that we are evaluating our algorithm on active regions that it has never seen before.

\subsection{Metrics}
\label{subsection:metrics}
There are many different metrics to assess the performance of a classification algorithm. These metrics are defined using four quantities: false positives (FP), false negatives (FN), true positives (TP), and true negatives (TN). The performance of a model depends not only on accuracy, but also on the cost of being wrong. Is it better to predict a flare and be wrong (a false positive) or to miss a big flare (a false negative)? The cost of an incorrect prediction depends on many application-specific factors. 

\begin{figure*}
\renewcommand{\tabcolsep}{0.02\textwidth}
\begin{tabular}{cc}
\includegraphics[width=0.28\textwidth]{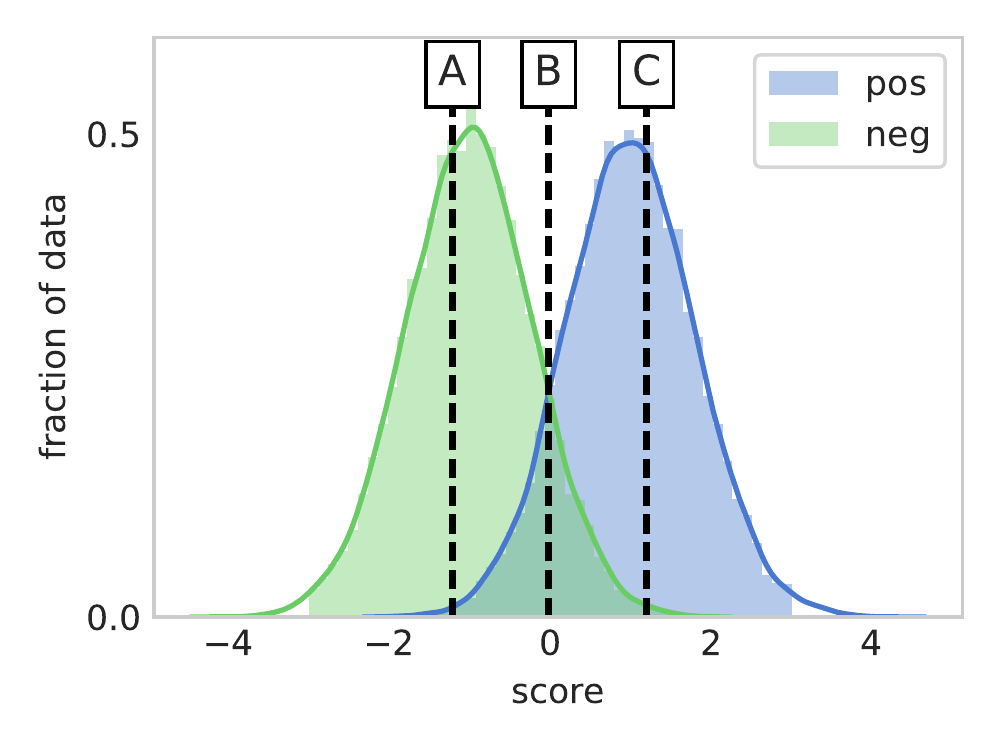} &
\includegraphics[width=0.70\textwidth]{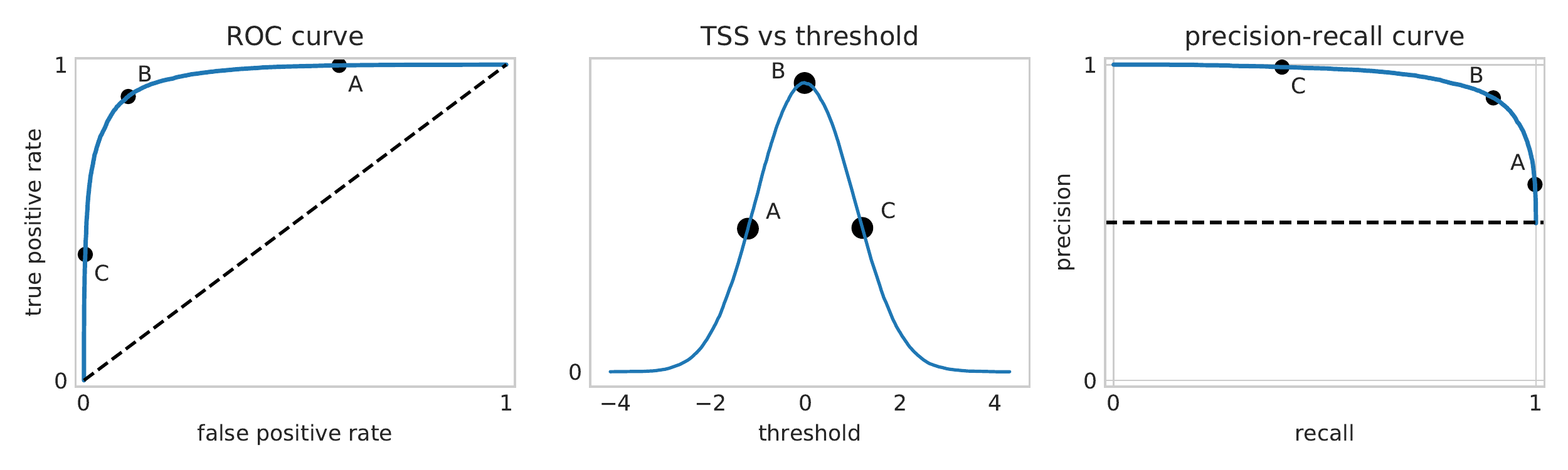} \\
\end{tabular}
\caption{Our binary classification tasks generate a distribution of score values for each class, positive and negative. We can see how varying that threshold (example thresholds A, B, and C) traces out several different curves. The Receiver Operating Characteristic (ROC) curve plots the true positive versus false positive rate as one varies this threshold. The TSS curve computes the True Skill Statistic as a function of the threshold. The precision-recall curve traces out the precision and recall as this threshold is varied. Note that there exists a threshold, B, which gives peak TSS. In this paper we focus on peak TSS and the area under the precision-recall curve.}
\label{fig:metrics_pedagogy}
\end{figure*}

Many classification algorithms produce a continuous value, or score, rather than a discrete binary outcome. To convert from these scores to a binary prediction, a threshold $\chi$ must be chosen -- events with scores above $\chi$ are classified as positive events, and those below $\chi$ are classified as negative. As the threshold varies, the fraction of true positives (the True Positive Rate, TPR) and false positives (FPR) varies, giving rise to the Receiver Operating Characteristic (ROC) curve (Figure~\ref{fig:metrics_pedagogy}). Different points on this curve represent trade-offs between a high number of false positives and a high number of false negatives, and the optimal threshold is application-dependent. As the optimal TPR/FPR trade-off (and thus score threshold) varies per application, we often use the total area under the ROC curve as a scalar that measures the performance of the algorithm. 

For problems with a high class imbalance, the area under the precision-recall curve can be more demonstrative of real-world performance. The precision of a classifier is the fraction of true positives out of all the correctly predicted events, i.e. TP/(TP+TN). In other words, a precision of 1.0 means that all the events were true positives -- that is, there are no false positives. The recall measures the fraction of true positives classified as positive, i.e. TP/(TP+FN). Perfect recall means all true positives were labeled positive -- that is, there are no false negatives.

However, precision and recall are sensitive to the class imbalance ratio, as are several other metrics (see \citealt{Bloomfield2012} and \citealt{Bobra2015} for a thorough discussion). The True Skill Statistic (TSS), also known as a Hansen-Kuipers skill score or the Peirce skill score \citep{Woodcock1976}, is not. As such, it is our metric of choice. The TSS is defined as follows:

\begin{equation}
\textrm{TSS} = \frac{\textrm{TP} \times \textrm{TN} - \textrm{FP} \times \textrm{FN}}{\textrm{P} \times \textrm{N}} = \textrm{recall} - \textrm{FPR}.
\end{equation}

The TSS ranges over $[-1, 1]$. A TSS of $1$ means all events were predicted correctly, $-1$ means all events were predicted incorrectly, and $0$ represents a random prediction. Note that the TSS is a function of the threshold $\chi$. As such, a range in choice of $\chi$ results in a range of possible TSS values.

\section{Algorithm}
\label{section:algorithm}
\begin{figure}
\centering
\includegraphics[width=6.7in]{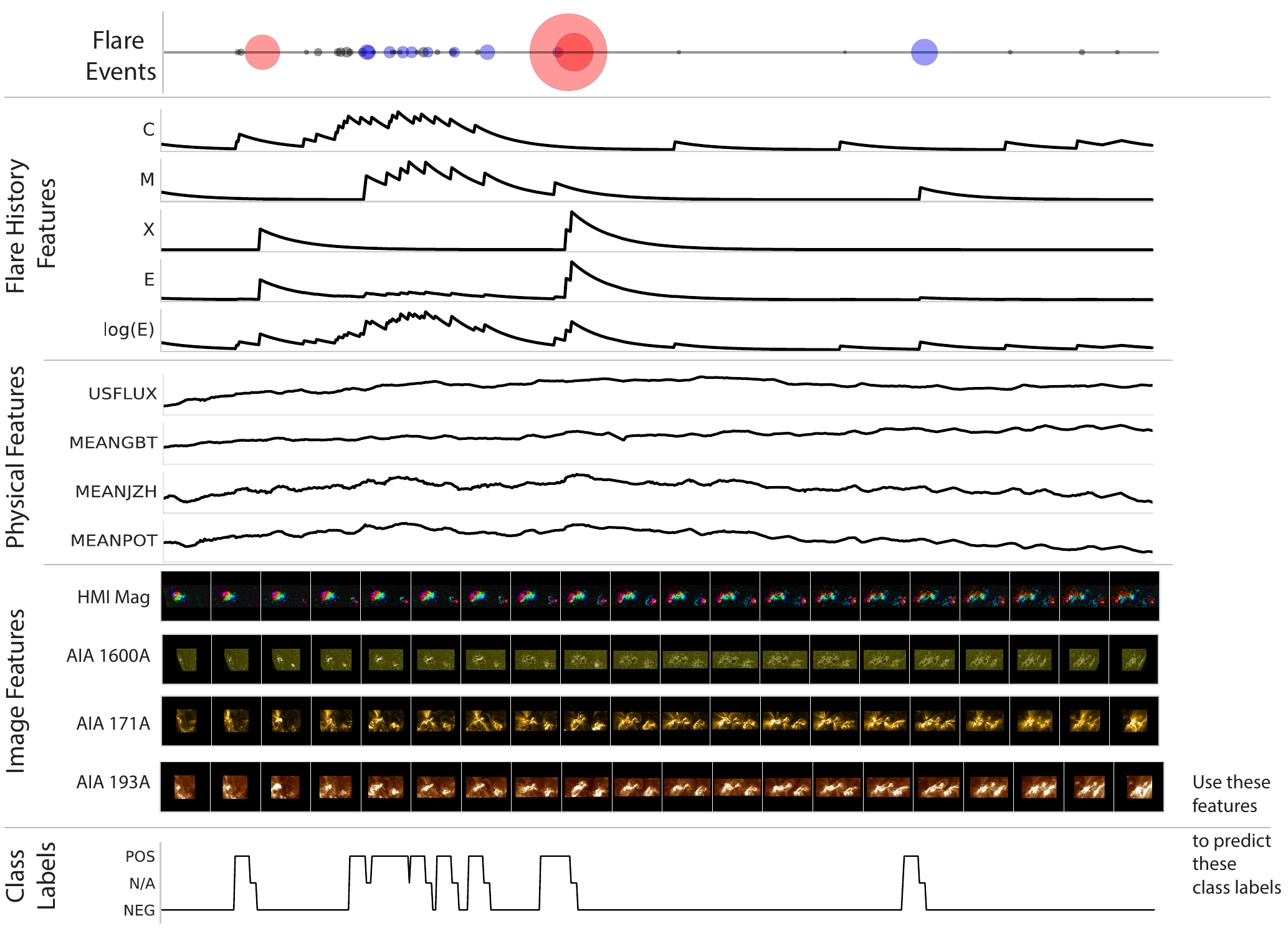} 
\caption{Overview of our features. We are interested in learning the relationship between various features of our input data and flaring activity. We use physical features (four shown, see Section~\ref{section:features:physical}), features that describe an active region's flaring history (shown here with a $\tau=6$h decay, see Section~\ref{section:features:flarehist}), and features automatically derived from HMI vector magnetogram and AIA UV and EUV image data (see Section \ref{section:features:imagefeatures}). X, M, and C-class flares are labelled at the top for reference. This plot shows features for SHARP 1449, which corresponds to NOAA Active Region 11429, from March 4, 2012 at 13:48:00 TAI to March 10, 2012 at 13:48:00 TAI.}
% from "feature timeseries plotting.ipynb"
\label{fig:all_data_example}
\end{figure}

Machine learning techniques can be broken down into a \textit{featurization} phase, which takes raw or preprocessed input data and extracts out relevant information, and a \textit{learning} phase, which uses the resulting collection of features to train a model. Thus, associated with each time-point $t$, we have both a feature vector $\vect{x}_t$ as well as an associated label $y_t$. Our goal is to learn from the training set of this data a function, $y_t = f(\vect{x}_t)$. We explain how we derive of our features, which are illustrated in Figure \ref{fig:all_data_example}, below. 

\subsection{Featurization}
\subsubsection{Physical Features}
\label{section:features:physical}
Our first set of features are those described in Table \ref{tab:params} and derived in \cite{Bobra14}. These 25 features describe various properties of solar active regions known to correlate with flaring activity (\citealt{lb03}, \citealt{fisher12}). These features are calculated every 12 minutes from the photospheric vector magnetic field image data (described in Section \ref{subsection:image_data}) for each active region throughout its lifetime.  Since these features have wildly differing dynamic ranges and are rejected for some time periods, we perform median imputation on all missing data before normalizing each feature to a zero-mean unit variance.

%%%%%%%%%%%%%%%% TABLE OF FORMULAE %%%%%%%%%%%%%%%%%%%

\begin{deluxetable*}{lll}
\tabletypesize{\scriptsize}
\tablecaption{Physical features describing solar active regions. \label{tab:physical}}
\tablewidth{0pt}
\tablehead{\colhead{Keyword} & \colhead{Description} &  \colhead{Formula} }
\startdata
{\sc totusjh} & Total unsigned current helicity  & ${H_{c_{total}}} \propto \sum |B_z \cdot J_z|$ \\
{\sc totbsq}  & Total magnitude of Lorentz force & $F  \propto \sum B^{2} $ \\
{\sc totpot} & Total photospheric magnetic free energy density & $ \rho_{tot} \propto  \sum \left( \vec{\textit{\textbf B}}^{\rm Obs} - \vec{\textit{\textbf B}}^{\rm Pot} \right)^2 dA $ \\
{\sc totusjz} & Total unsigned vertical current & ${J_{z_{total}}} =  \sum |J_{z}|dA$ \\
{\sc absnjzh} & Absolute value of the net current helicity & ${H_{c_{abs}}} \propto \left| \sum B_z \cdot J_z \right|$ \\
{\sc savncpp} & Sum of the modulus of the net current per polarity & $J_{z_{sum}} \propto \Big\vert \displaystyle\sum\limits^{B{_z^+}} J{_z}dA \Big\vert + \Big\vert \displaystyle\sum\limits^{B{_z^-}} J{_z}dA \Big\vert $ \\
{\sc usflux} & Total unsigned flux & $\Phi = \sum|B_{z}|dA$ \\
{\sc area\_acr} &  Area of strong field pixels in the active region &  Area $ = \sum$ Pixels \\
{\sc totfz}  &  Sum of z-component of Lorentz force & $F_{z}  \propto \sum (B_{x}^{2} + B_{y}^{2} - B_{z}^{2}) dA$ \\
{\sc meanpot} & Mean photospheric magnetic free energy & $ \overline{\rho} \propto \frac{1}{N} \sum \left( \vec{\textit{\textbf B}}^{\rm Obs} - \vec{\textit{\textbf B}}^{\rm Pot} \right)^2 $ \\
{\sc r\_value} &  Sum of flux near polarity inversion line &  $\Phi = \sum|B_{LoS}|dA$ within R mask \\
{\sc epsz} & Sum of z-component of normalized Lorentz force  & $\delta F_{z}  \propto \frac{\sum (B_{x}^{2} + B_{y}^{2} - B_{z}^{2})}{ \sum B^{2}}$  \\
{\sc shrgt45} & Fraction of Area with Shear $> 45^\circ$  & Area with Shear $>45^\circ$ / Total Area \\
{\sc meanshr} & Mean shear angle  & $ \overline{\Gamma} = \frac{1}{N} \sum \arccos \left( \frac{\vec{\textit{\textbf B}}^{\rm Obs} \cdot \vec{\textit{\textbf B}}^{\rm Pot}}{|B^{\rm Obs}|\,|B^{\rm Pot}|} \right)$ \\
{\sc meangam} & Mean angle of field from radial &  $\overline{\gamma} = \frac{1}{N} \sum \arctan\left(\frac{B_h}{B_z}\right)$  \\
{\sc meangbt} & Mean gradient of total field & $\overline{\left|{\nabla B_{\rm tot}}\right|} = \frac{1}{N} \sum \sqrt{\left(\frac{\partial B}{\partial x}\right)^2 + \left(\frac{\partial B}{\partial y}\right)^2}$ \\
{\sc meangbz} & Mean gradient of vertical field & $\overline{\left|{\nabla B_z}\right|} = \frac{1}{N} \sum \sqrt{\left(\frac{\partial B_z}{\partial x}\right)^2 + \left(\frac{\partial B_z}{\partial y}\right)^2}$ \\
{\sc meangbh} & Mean gradient of horizontal field & $\overline{\left|{\nabla B_h}\right|} = \frac{1}{N} \sum \sqrt{\left(\frac{\partial B_h}{\partial x}\right)^2 + \left(\frac{\partial B_h}{\partial y}\right)^2}$ \\
{\sc meanjzh} & Mean current helicity ($B_{z}$ contribution) & $\overline{H_c} \propto \frac{1}{N} \sum B_z \cdot J_z $ \\
{\sc totfy}  &  Sum of y-component of Lorentz force & $F_{y} \propto \sum B_{y} B_{z} dA$ \\
{\sc meanjzd} & Mean vertical current density & $\overline{J_z} \propto \frac{1}{N} \sum \left(\frac{\partial B_y}{\partial x} - \frac{\partial B_x}{\partial y}\right) $ \\
{\sc meanalp} & Mean characteristic twist parameter, $\alpha$ & $ \alpha_{total} \propto \frac{\sum J{_z} \cdot B_z}{\sum B{^2_z}} $ \\
{\sc totfx} &  Sum of x-component of Lorentz force & $F_{x} \propto - \sum B_{x} B_{z} dA$ \\
{\sc epsy}  & Sum of y-component of normalized Lorentz force & $\delta F_{y}  \propto \frac{-\sum B_{y} B_{z}}{ \sum B^{2}}$ \\
{\sc epsx}  & Sum of x-component of normalized Lorentz force & $\delta F_{x}  \propto \frac{\sum B_{x} B_{z}}{ \sum B^{2}}$ \\
\enddata
\tablecomments{The {\it Keyword} column indicates the name of the FITS header keyword in the HMI  SHARP vector magnetic field data. The {\it Description} column indicates the computed physical quantity per the equation indicated in the {\it Formula} column. See \citet{Bobra14} for more detail about the image data and keyword calculation.\label{tab:params}}
\end{deluxetable*}

%%%%%%%%%%%%%%%% TABLE OF FORMULAE %%%%%%%%%%%%%%%%%%%

\subsubsection{Flare History Features}
\label{section:features:flarehist}

\begin{figure}
\centering
\includegraphics[width=1.5in]{flarehist_explanation.ai} 
\includegraphics[width=3.5in]{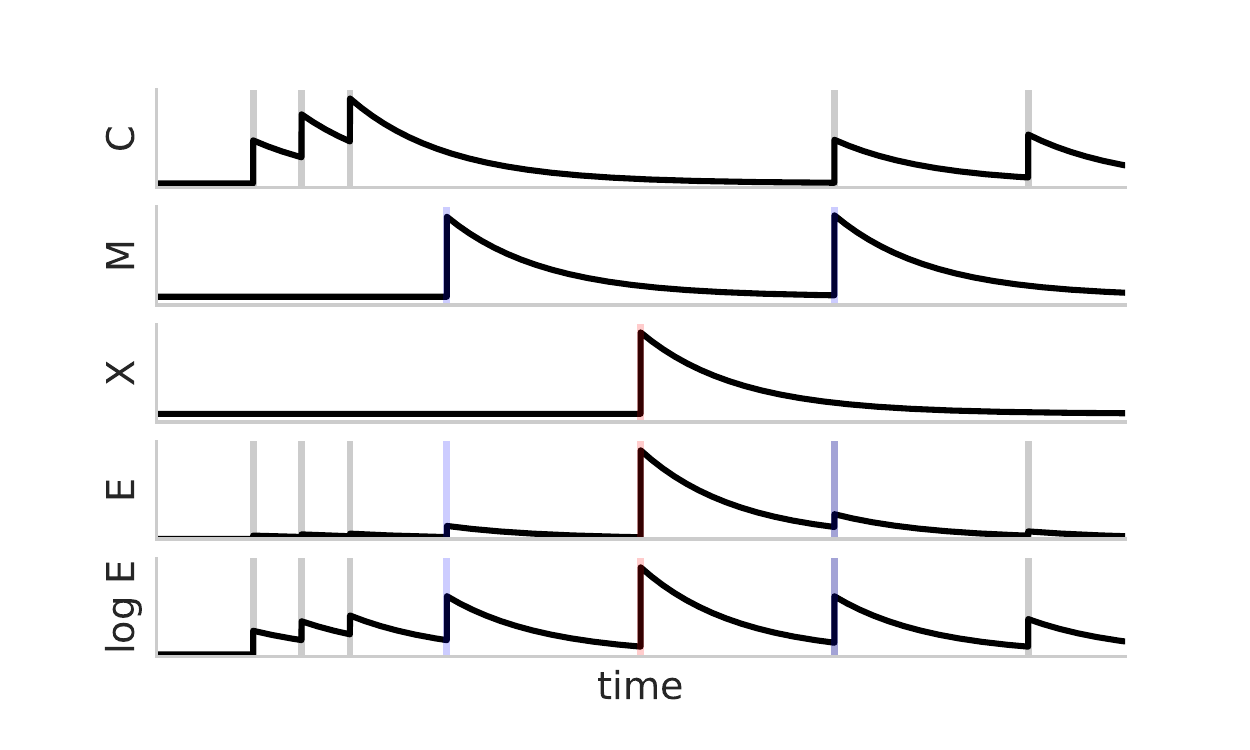}
\caption{\textbf{Flare history features} : For each SHARP we construct a timeline of flaring history, and apply a weighted decaying exponential with decay $\tau$ for each flare peak. We do this for timeseries consisting of the individual flare classes as well as the total flare energy and the log of this energy. We construct this 5-dimensional timeseries for a variety of different decay constants $\tau$.  }
\label{fig:flare_history}
\end{figure}

Knowledge of whether an active region flared in its past can greatly improve the ability to forecast its future (\citealt{falconer12}, \citealt{Wheatland2004}). As such, we derived a number of features that capture the flaring history of an active region. For each active region we take the list of associated flares in the GOES solar flare catalog and construct the following time series:
\begin{equation}
f(t) = \sum_{i \in F_c} E_i\delta(t-t_i)
\end{equation}
where $F_c$ are the flares of a given class for that SHARP and $t_i$ is the time of the indicated flare with intensity $E_i$ (see Figure \ref{fig:flare_history}). We create five of these flare time series. First, a time-series of only $C$ class flares, where we keep $E_i=1$. This is thus simply a time series of $C$-class flare occurrences. We then create a similar series for $M$ and $X$-class flares. The final two time series contain all observed flares regardless of class, but scaled by setting $E_i$ to the observed flare magnitude or log-magnitude. These time-series are then convolved with exponentially-decaying windows $\tau$ of varying length.

\subsubsection{Image Features}
\label{section:features:imagefeatures}

\begin{figure*}
\centering
\includegraphics[width=\linewidth]{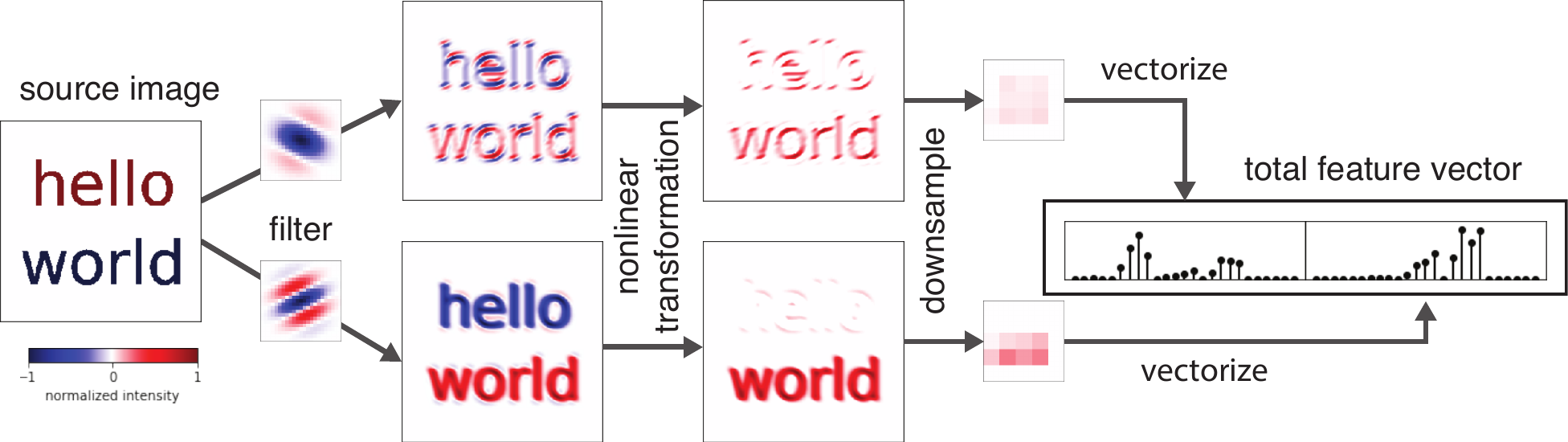} 
\caption{Featurization for image data involves taking an input image and convolving it with a number of pre-selected filters (two are shown here). The resulting intermediate images are transformed by a per-pixel non-linearity and then downsampled. The resulting downsampled images are vectorized and concatenated into a total feature vector.}
\label{fig:image_featurize}
\end{figure*}

We automatically derive features from the HMI and AIA image data via a parametrized nonlinear filtering process that ultimately results in a large real-valued feature vector per image. The fundamental featurization operation is a convolution followed by a per-pixel nonlinear transformation and then downsampling. In other words, the featurization process consists of three steps. Step 1: An input $n \times n \times c$ image $x_i$ is convolved with a $k \times k \times c$ filter $h_j$. Step 2: The resulting filtered image is passed through a per-pixel nonlinearity. Step 3: The resulting field is downsampled, or pooled, to a smaller $m \times m$ image. These $m^2$ pixels are then concatenated into the feature vector for this image. If we process an image with $F$ total features we end up with a feature vector consisting of $F\cdot m^2$ elements. The following is a basic outline of these three steps (see Figure~\ref{fig:image_featurize} for a schematic example).

\begin{figure}
\centering
\includegraphics[width=4in]{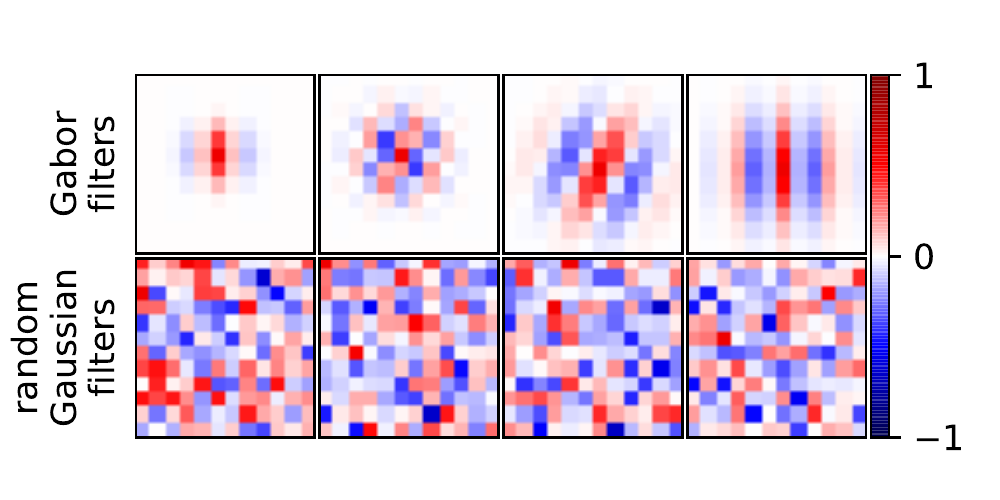} 
\caption{Example Gabor filters (top) and random Gaussian filters (bottom) used to featurize
the input images.}
\label{fig:example_filters}
\end{figure}

\begin{enumerate}
\item For our convolution step we consider two classes of filters: Gabor filters and random filters (see Figure \ref{fig:example_filters}). Gabor filters \citep{Gabor1947} have been used extensively for image featurization (e.g. \citealt{Kamarainen2006}). A Gabor filter is a function of space $g(x, y)$ which consists of a spatial sinusoid modulated by a Gaussian envelope. The entire filter is rotated at an angle $\theta$. Letting
$x' = x \cos \theta + y \sin \theta$, $y' = -x \sin \theta + y \cos \theta$ represent the filter in the
$\theta$-rotated coordinate frame, the filter is
\begin{equation}
g(x, y)  = \exp\Big(-\frac{x'^2 + \gamma^2y'^2}{2\sigma^2}\Big)\sin\Big(2\pi \frac{x'}{\lambda}
 + \psi \Big)
\end{equation}

where $\lambda$ controls the period of the sinusoid, $\theta$ is the rotation of the coordinate axis, $\psi$ is the phase offset, $\sigma$ is the standard deviation of the Gaussian envelope and $\gamma$ controls the ellipticity. 

Random filters may appear counter-intuitive for filtering applications such as ours, but there are theoretical reasons why they can be effective.

A desirable property of a image featurization is spatial-invariance: we would like two images with similar characteristics at different pixel locations (e.g top-left corner and bottom-right corner), to be close in our featurized space. For example, two images of the same sunspot at different times will have the same characteristics (the sunspot) at different locations as it moves across the sun. We can formalize this concept a bit further by defining an inner product between multi-channel images that exhibits this property.

Let $\mathcal{P}_{k}(x)$ be the set of all $k \times k$ sub-images (square patches) in an multi-channel image $x$. Let $d$ be any distance function that measures the scaled dissimilarity between two patches (i.e., the maximum distance between patches is $1.0$).

\begin{equation}
\kappa(x, y) = \displaystyle\sum_{x_{i} \in \mathcal{P}(x)} \displaystyle\sum_{y_{i} \in \mathcal{P}(y)} (1 - d(x_{i}, y_{i}))
\label{comparison}
\end{equation}

Since Equation ~\ref{comparison} compares all $k \times k$ sub patches across all spatial locations, the feature space induced by this inner product exhibits a degree of spatial-invariance. We can then construct a featurization $F(x)$ of our images where the \textit{Euclidean} inner product between two feature vectors $\langle F(x_1), F(x_2) \rangle \approx \kappa(x_1, x_2)$. In fact, we can approximate this for common non-linear distance functions (\citealt{mairal2014convolutional}, \citealt{cho2009kernel}, \citealt{rahimi2008random}) by convolving, rectifying (see the next step), and downsampling each image with a set of random filters. We chose to approximate the arc-cosine distance function of \cite{cho2009kernel} due its expressive nature and ease of approximation.

We thus convolve each input image with a series of random filters where each pixel in each filter is drawn from a zero-mean normal distribution with a fixed standard deviation, also known as the bandwidth. As we increase the number of filters, we better approximate this function -- but too many filters may lead to overfitting.

\item We then apply a symmetric nonlinear rectifier to each resulting filtered pixel: 
 \[
h(x) =
  \begin{bmatrix}
    \max(x-b, 0)\\
    \max(-(x-b),0)
  \end{bmatrix}
\]
where $b$ is the bias term. Without this nonlinearity, our method would be entirely linear and thus unable to learn nonlinear relationships. 

\item We then downsample the resulting filtered images to $1\times 1$, i.e. a single scalar. This effectively results in measuring the total energy present in a given filter. 

\end{enumerate}

\begin{figure}
\centering
\includegraphics[width=4in]{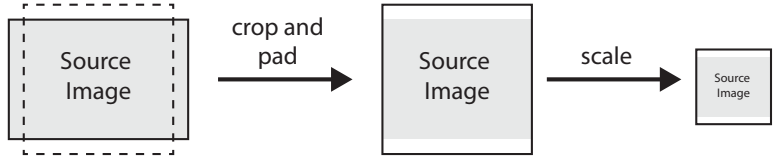} 
\caption{Image preprocessing for HMI and AIA images places the original image on a canvas, cropping and padding as necessary to achieve a consistent square aspect ratio. Images are then downsampled to the final $256\times 256$ pixels.}
\label{fig:image_preprocess}
\end{figure}

We assemble the HMI vector magnetograms into images by first creating a single 3-channel image containing $B_r$, $B_\phi$, and $B_\theta$, which, as mentioned in Figure \ref{fig:example}, are remapped from CCD coordinates to heliographic Cylindrical Equal-Area coordinates. We then augment each image with six additional channels: $|B_r|$, $|B_\phi|$, and $|B_\theta|$, as well as $\sqrt{|B_r \cdot B_\phi|}$, $\sqrt{|B_r \cdot B_\theta|}$, and $\sqrt{|B_\theta \cdot B_\phi|}$, in an attempt to capture per-channel nonlinear interactions, which are known to be important in the physical features. We use the bitmap arrays produced by the HMI active region detection module and disambiguation module (see the last two rows in Table \ref{tab:datavolume}) to mask out regions of the image with low signal-to-noise, setting them to zero. To account for the wide variance in SHARP sizes, we center each SHARP image on a canvas, cropping and padding as necessary. The resulting square image is then resized to $256\times 256$ pixels (Figure~\ref{fig:image_preprocess}). We evaluated both a wide variety of Gabor filters and random filters for the HMI 9-channel magnetogram data, choosing a series of 1024 $14\times14$ filters with a bandwidth of $0.1$ and a bias of $100$. 

We then select the full-disk AIA 171$\AA$, $193\AA$, and $1600\AA$ images closest in time to each HMI record. We use the coordinates provided by the HMI SHARP partial-disk image to select the equivalent region in the AIA image. Note that we retain the AIA data in CCD coordinates as it is difficult to transform to Cylindrical Equal-Area coordinates without knowing the height of the observed UV and EUV emission. We treat each AIA channel independently, producing a separate feature vector for each wavelength. We chose 1024 random filters with a bandwidth of $0.1$ for all three channels. For the AIA $171\AA$ channel, we found that $6$-pixel patches with a bias of $50$ performed best; we found $12$ and $6$-pixel patches with a bias of $50$ and $100$ performed best for the AIA $193\AA$ and $1600\AA$ channels, respectively.

\subsection{Learning}
\label{section:Learning}
\begin{figure}
\centering
\includegraphics[width=2.6in]{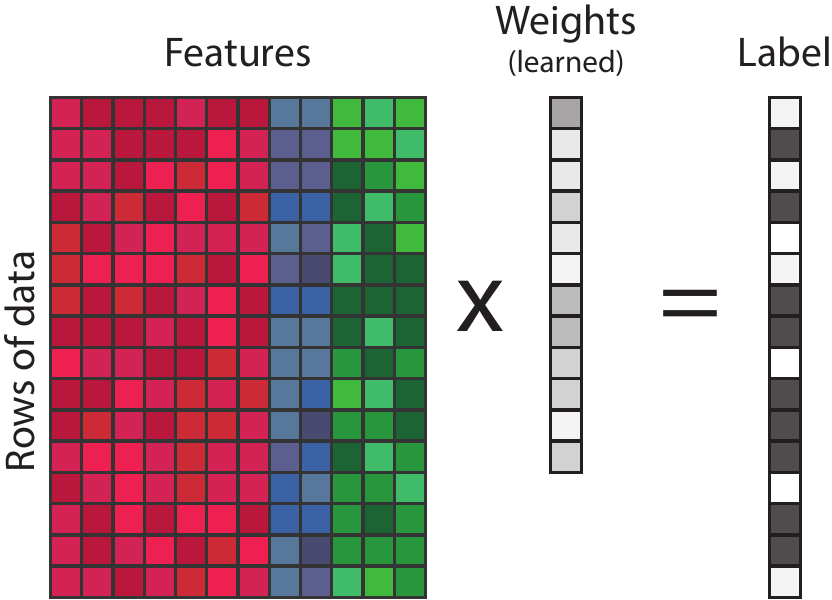} 
\caption{All of our models ultimately depend on linear classifiers which compute different types of features for each datapoint, and take the dot product of that feature vector with a learned weight vector to produce a prediction. }
\label{fig:linear_model}
\end{figure}

In principle, any classifier, ranging from simple linear classifiers to complex methods like neural networks and random forests \citep{Bishop2006}, could be used on a featurized dataset. We focus on on linear classifiers, which model the output $y_i \in \{-1, 1\}$ as a linear function of the input features
\begin{equation}
y_i = \vect{w}^T \vect{x_i}
\end{equation}
where $\vect{x_i}$ is a vector of features for datapoint $i$ and $\vect{w}$ is the vector of model parameters learned (Figure \ref{fig:linear_model}). We care about the model's ability to predict future flaring events on unseen data, and it may be the case that our model fits the training data too accurately, at the expense of future predictive performance. This is termed overfitting in the machine learning literature \citep{Bishop2006}, and we compensate for it by preferring models which are simpler, a process known as regularization. 

%\fixme{in the results section, we talk about using least squares, Lasso, and a fast approximation to the RBF-kernel SVM. so we need to mention exactly what these models are in the learning section. they don't have to be extensive about it (we can refer the readers to other papers) but we need to be explicit about how many models we used and how they differ from one another}

If $\vect{X} \in \Reals^{M \times N}$ is a matrix of $M$ rows of featurized data (see Figure~\ref{fig:linear_model}), then we can solve the following optimization problem:
\begin{equation}
\min_{\vect{w}} ||\vect{X}\vect{w} - \vect{y}||_2^2 +\lambda ||\vect{w}||_p
\end{equation}
where $\lambda$ is the regularization parameter, controlling the trade-off between the data fit term $||\vect{X}\vect{w} - \vect{y}||_2^2$ and the regularization term $ ||\vect{w}||_p$. This term penalizes weight vectors $\vect{w}$ with large norm.
When $p=2$ this problem is similar to Tikhinov-regularized linear regression (which uses $||w||_2^2$), and when $p=1$ this problem favors sparse solutions, in which most $w_i=0$ and thus don't contribute to the model. This sparse regression model goes by many names, including the Lasso, and allows us to determine which features contribute most to a particular model's predictive performance. When $p=2$, we can quickly solve the problem by direct inversion, as its runtime grows with the number of features, not the amount of data. This enables rapid evaluation of a large number of folds for cross-validation. We will henceforth refer to the $p=2$ case as our "linear classifier" and the $p=1$ case as our "sparse linear classifier" to avoid confusion. 

Our dataset thus contained 1,256,700 time steps, comprising 1,256,700 HMI SHARP images and 470,951 AIA full-disk images cut into $\sim$3,600,000 partial-disk images to correspond with the HMI SHARP regions (see Table~\ref{tab:datasize}). To cope with processing this volume of data, which comprises 5.5 TB in total, we developed PyWren: a distributed computing framework optimized for cloud-based image processing applications such as ours \citep{Jonas2017}. The scale and use of cloud compute infrastructure enabled us to featurize all of the AIA and HMI image data for a given filter configuration in tens of minutes. 

While more complex problem formulations, featurizations, and models are indeed possible (See Section~\ref{section:conclusion} for future directions), here we focus primarily on entirely feed-forward image featurization and linear classifiers. This is partly due to the scale of the data -- training more complex models takes a long time, and we wanted to rapidly evaluate the impact of changes in our pre-processing steps. This is also due to our insistence on careful cross-validation (of 100 separate training and testing folds) to accurately gauge the uncertainty in and sensitivity of our models. 

\begin{deluxetable}{lllrr}
\tabletypesize{\footnotesize}
\tablecaption{Data Type and Volume \label{tab:datasize}}
\tablehead{
\colhead{Type} & \colhead{Series} & \colhead{Segment} &  \colhead{Record Count} &  \colhead{Volume} }
\startdata
AIA 1600 \AA	   & \sdoseries{aia.lev1} & \segment{image\_lev1}		& 156,933		& 1.64 TB\\
AIA 171 \AA		   & \sdoseries{aia.lev1} & \segment{image\_lev1}		& 156,992		& 1.93 TB \\
AIA 193 \AA		   & \sdoseries{aia.lev1} & \segment{image\_lev1}		& 156,966 		& 1.94 TB \\
HMI	B${_r}$	 	   & \sdoseries{hmi.sharp\_720s\_cea} & \segment{Br}				& 1,256,700 	& 478 GB \\
HMI	B${_{\phi}}$   & \sdoseries{hmi.sharp\_720s\_cea} & \segment{Bp}				& 1,256,700		& 473 GB\\
HMI	B${_{\theta}}$ & \sdoseries{hmi.sharp\_720s\_cea} & \segment{Bt}				& 1,256,700 	& 490 GB \\
HMI HARP		   & \sdoseries{hmi.sharp\_720s\_cea} & \segment{bitmap}			& 1,256,700		& 272 GB\\
HMI	Disambiguation & \sdoseries{hmi.sharp\_720s\_cea} & \segment{conf\_disambig}	& 1,256,700		& 48 GB \\
\enddata
\tablecomments{SDO HMI and AIA data are cataloged into various data series and can be downloaded from the Joint Science Operations Center at \url{http://jsoc.stanford.edu}. These series contain image data, or segments, and metadata that are merged into FITS files upon export. The {\it Type} column indicates the type of data, the {\it Series} column indicates the name of the data series, the {\it Segment} column indicates the name of the segment, or associated data array, the {\it Record Count} indicates the number of unique records, and {\it Volume} is the total size of those records. Since we use an HMI data series that contains partial-disk images of active regions, and many active regions can appear on the disk at the same time, this series may contain multiple records for any given time. Since we use an AIA data series that contains full-disk imagery, it only contains one image for any given time. As such, there are fewer unique input AIA images than HMI images in our dataset.\label{tab:datavolume}} 
\end{deluxetable}

\section{Results}
\label{section:results}
We evaluate the predictive performance for each of our features and then combine these to evaluate their aggregate performance. In all cases, we focus on the two prediction tasks ($T$ = 2 and and 24 hours) and two performance metrics (area under the precision-recall curve and TSS). These metrics are displayed as box plots with median and interquartile ranges, as well as a scatter distribution of per-fold performance for the indicated metric. We use a linear classifier for all feature subsets; for features with interpretable features (physical and flare history) we also apply sparse linear classifiers in an attempt to ascertain which features are most useful. 

\subsection{Physical Features}

\begin{figure*}
\centering
\includegraphics[width=0.85\linewidth]{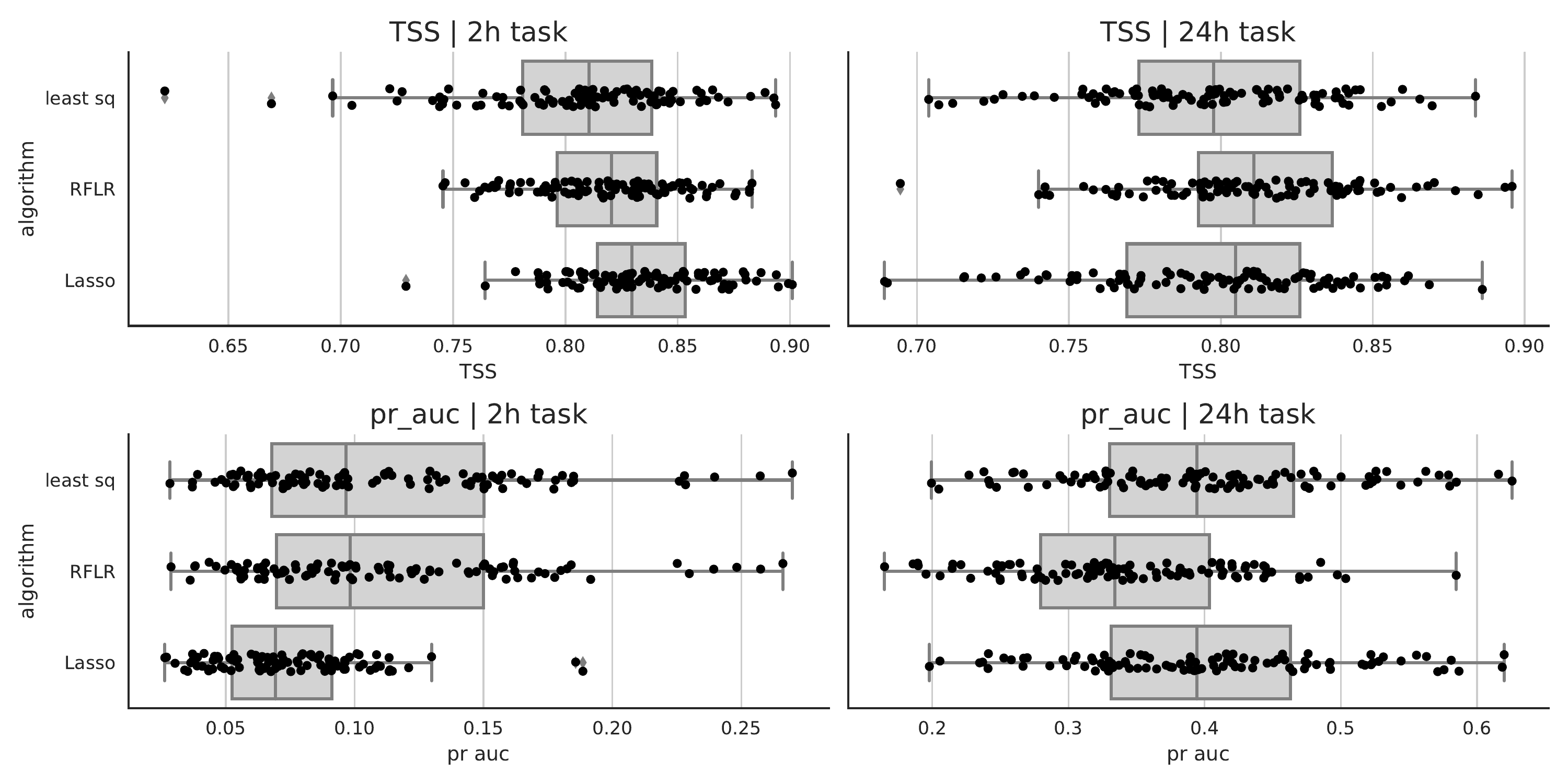} 
\caption{The performance, as measured by the TSS and area under the precision-recall curve (labeled as pr\_auc), of our baseline physical features used in conjunction with three different algorithms -- least-squares linear classifier (labeled as least sq), a l1-regularized sparse linear classifier (Lasso), and the RBF-kernel SVM (RFLR) -- to predict whether an active region will flare within $T$ = 2 and 24 hours. Our results are from 100-fold cross validation. We find that we can achieve similar results to those in \citet{Bobra2015}, who obtained a TSS of $0.761 \pm 0.039$ for a task similar to the $T$=24 hours case.} 
\label{fig:results_physical_metrics}
\end{figure*}

We began by trying to reproduce the results in \citet{Bobra2015}, who used the physical features described in Table \ref{tab:params} and a fast approximation to the Radial Basis Function (RBF) kernel Support Vector Machine (SVM) to predict whether an active region would flare in exactly $T$ = 24 hours. However, we predicted whether an active region would flare {\it within}  $T$ = 24 hours, which is a much more realistic problem. While our dataset spanned the same time period as \citet{Bobra2015}, we trained on significantly more data and used three different learning models. Figure \ref{fig:results_physical_metrics} shows that all three of our learning models perform equally well and we can reproduce the results in \cite{Bobra2015}, who obtained a TSS of $0.761 \pm 0.039$ for a T equal to exactly 24 hours (see figure~\ref{fig:prediction_task} for an explanation). 

\begin{figure}
\centering
\includegraphics[width=0.5\textwidth]{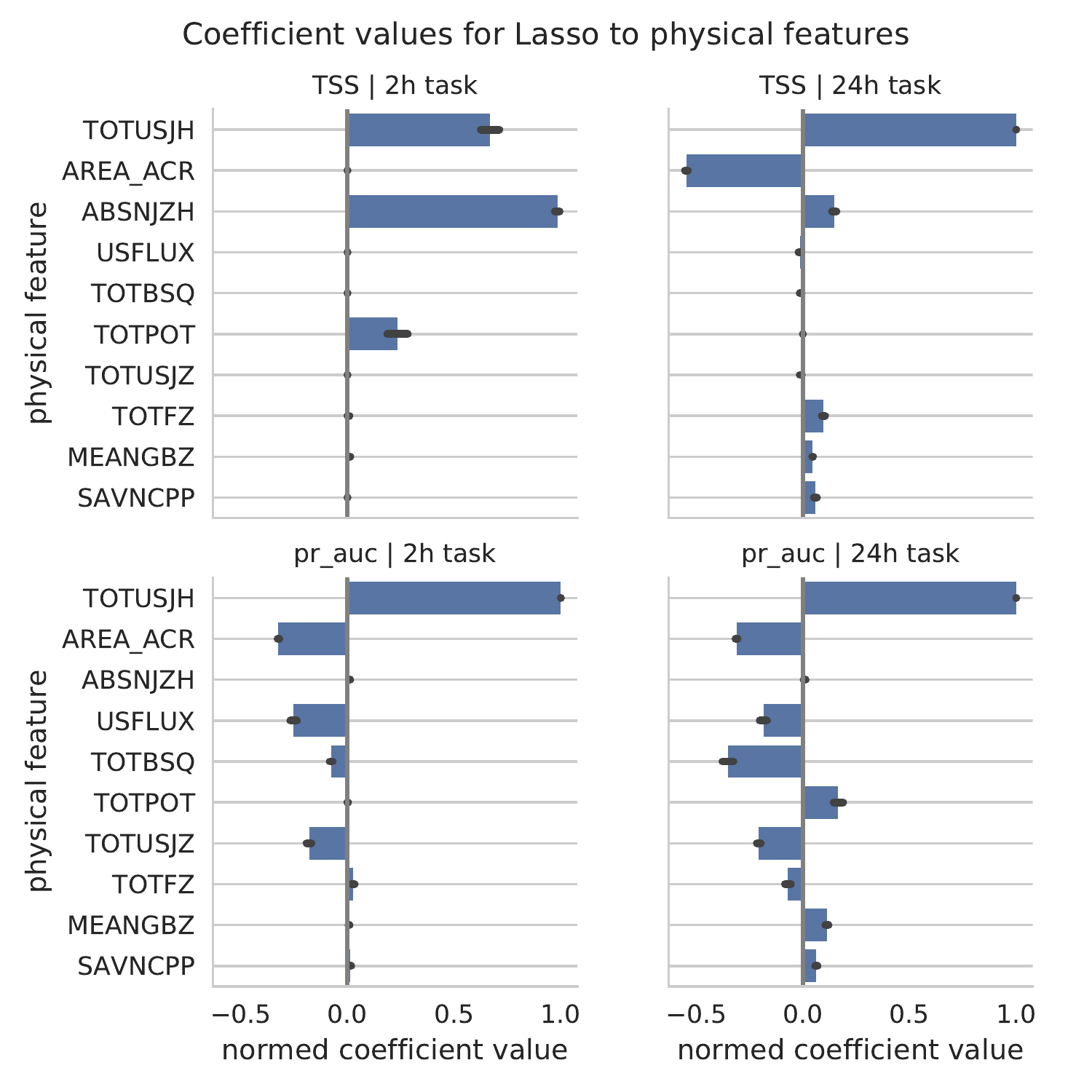} 
\caption{These four panels show the top 10 coefficients for the sparse linear classifier, or Lasso, model optimized for each task and metric using only the physical features as an input. Error bars indicate per-fold standard error. The total unsigned current helicity has the highest predictive capacity in both the $T$ = 2 and 24 hour tasks and for both metrics.}
\label{fig:results_physical_coeff}
\end{figure}

We can also evaluate the coefficient values returned by the sparse linear classifier to understand which features are most useful for our prediction task. For each task and metric combination we find the best-performing value of the regularization parameter $\lambda$ across all 100 folds, and then plot the distribution of coefficient values in Figure~\ref{fig:results_physical_coeff}. 

We find that the total unsigned current helicity has the highest predictive capacity in both the $T$ = 2 and 24 hour tasks, a result confirmed in previous studies (e.g. \citealt{leka07}). Because this quantity involves multiplying the vertical component of the current, $J_{z}$, with the vertical component of the magnetic field, $B_{z}$, current helicity identifies regions that contain twisted magnetic field. 

Three of the four panels in Figure ~\ref{fig:results_physical_coeff} also show an inverse relationship between current helicity and area. This implies that area in and of itself is not a sufficient predictor and only active regions that contain a large area and sufficient current helicity will flare. 

We also find that adding more physical features beyond a few does not seem to have a large impact, a result corroborated by \citet{leka07} and \citet{Bobra2015}. As such, this motivates adding different types of information in the form of flaring history and EUV and UV images. We will explore the results of these features in the subsequent sections.

\subsection{Flare History Features}
\begin{figure*}
\renewcommand{\tabcolsep}{0.0018\textwidth}
\centering
\includegraphics[width=0.9\textwidth]{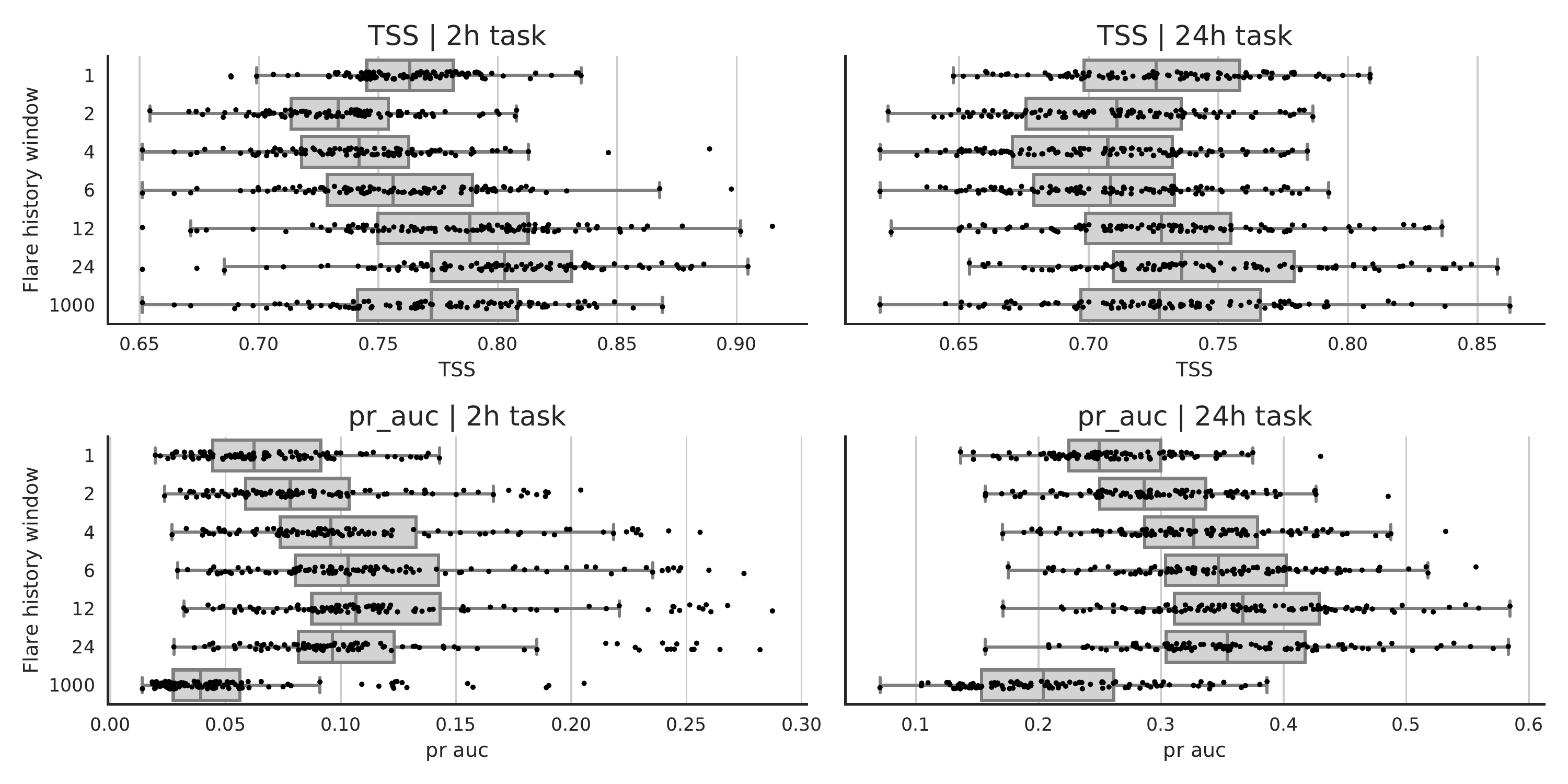} 
\caption{The performance, as measured by the TSS and area under the precision-recall curve, of our flaring history features used in conjunction with the sparse linear classifier, or Lasso, model across all tasks and metrics. In all cases, we find that values of the decay parameter $\tau$ equal to 12 and 24 hours perform the best.}
\label{fig:results_flarehist_metrics}
\end{figure*}

Figure \ref{fig:results_flarehist_metrics} shows the predictive performance of the five flaring history features across 1, 2, 4, 6, 12, 24, and 1,000 hour values for the decay parameter $\tau$. We find that prior flaring activity has significant predictive capacity in and of itself, a result corroborated by many others (e.g. \citealt{falconer12}). In all cases, we find that values of the decay parameter $\tau$ equal to 12 and 24 hours perform the best.

\begin{figure*}
\renewcommand{\tabcolsep}{0.0018\textwidth}
\centering
\includegraphics[width=0.8\textwidth]{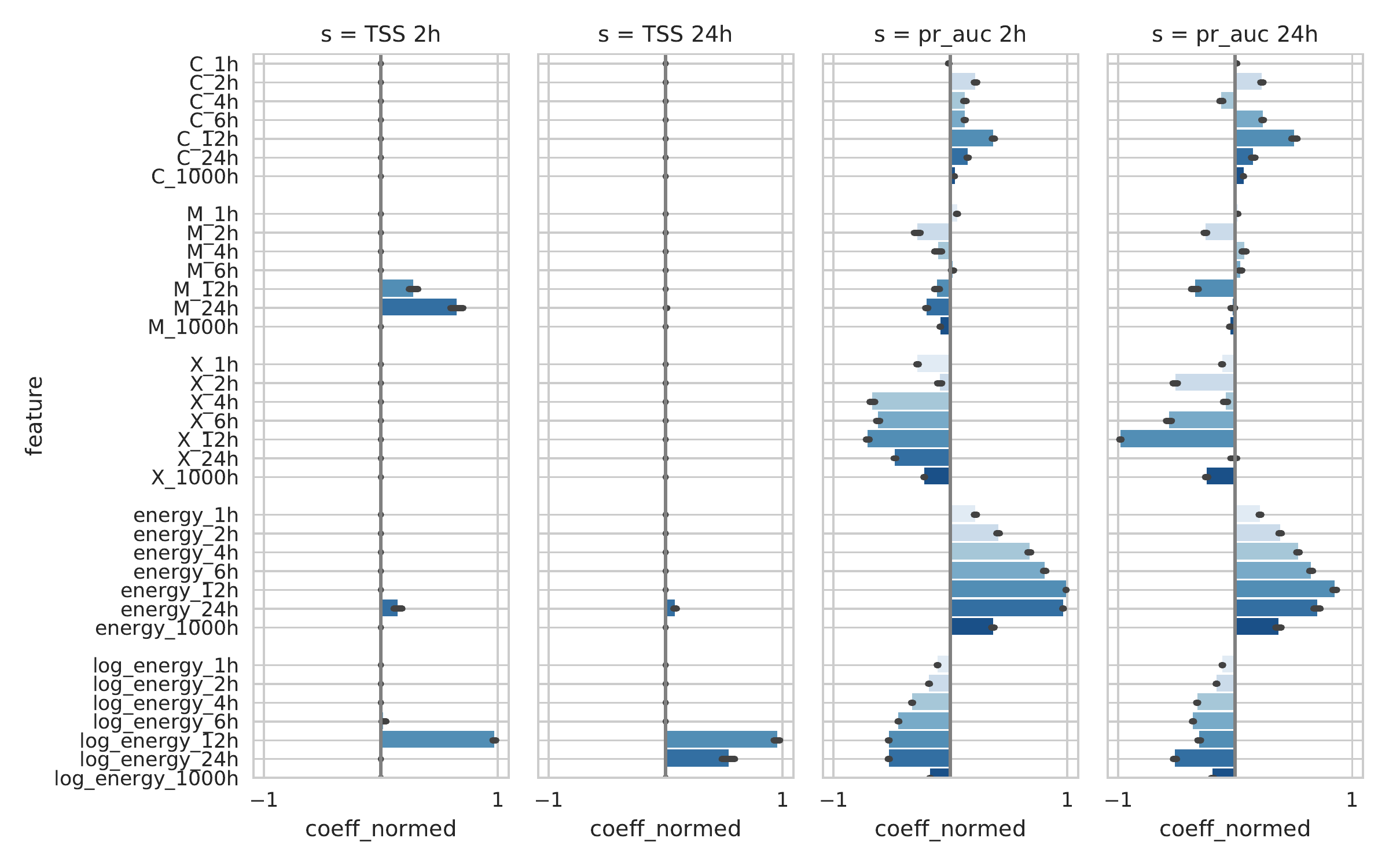} 
\caption{These four panels show the distribution of coefficients for the sparse linear classifier, or Lasso, model optimized for each task and metric using only the flaring history features as an input. Error bars indicate per-fold standard error. The energy expended by an active region, as measured by the observed flare magnitude or log-magnitude, 12 hours prior to a flare is the most useful predictor of whether it will flare again.}
\label{fig:results_flarehist_coeff}
\end{figure*}

As with the physical features, the coefficient values returned by the sparse linear classifier, or Lasso, model can help us understand which flaring history features are most useful for our prediction task. Figure \ref{fig:results_flarehist_coeff} shows that the energy expended by an active region, as measured by the observed flare magnitude or log-magnitude, 12 hours prior to a flare is the most useful predictor of whether it will flare again. Surprisingly, we found that this energy budget is not necessarily associated with previous large flares. This could be due to small-number statistics in our sample of X-class flares or, as \citet{falconer12} theorizes, ``simply because an active region's chance of having another major eruption increases with increasing time since its last major eruption."

\subsection{Combining Features}
Finally, we combine the hand-engineered physical and flaring history features with the features automatically derived from the HMI and AIA image data. We then use all these features in conjunction with a linear classifier.

\begin{figure*}
\renewcommand{\tabcolsep}{0.0018\textwidth}
\begin{tabular}{cc}
\includegraphics[angle=0,width=0.496\textwidth]{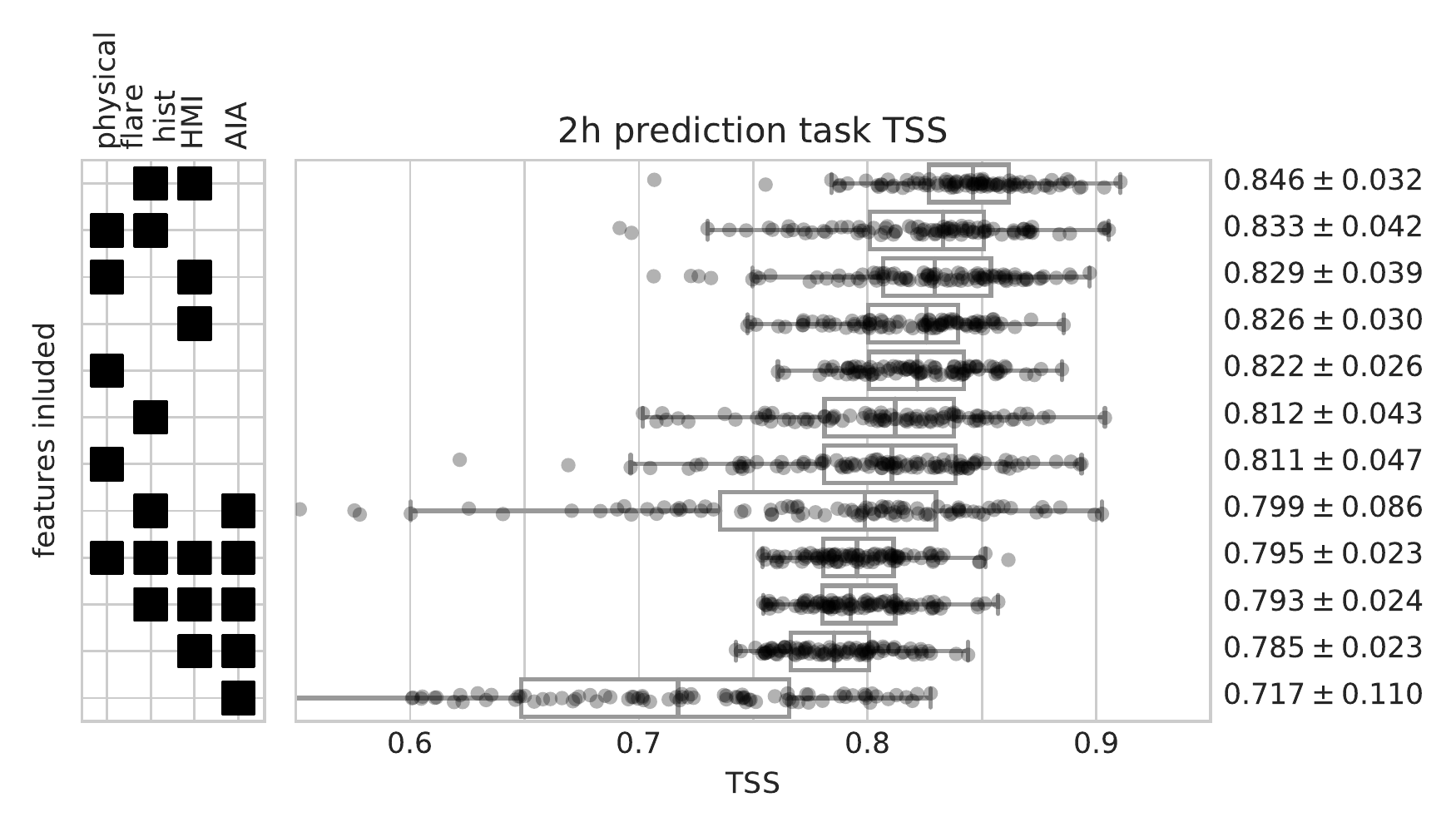} &
\includegraphics[angle=0,width=0.496\textwidth]{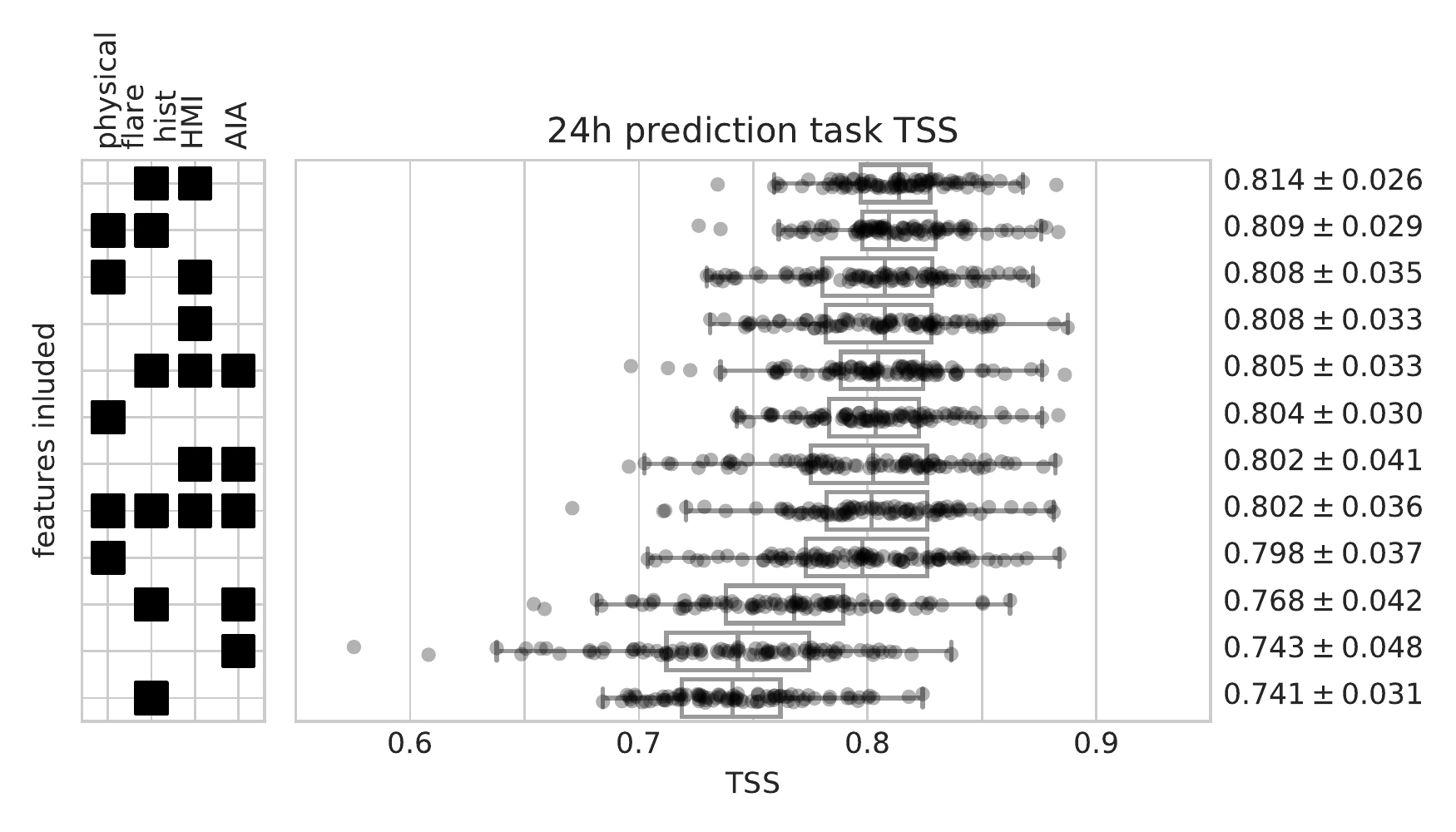}\\
\includegraphics[angle=0,width=0.496\textwidth]{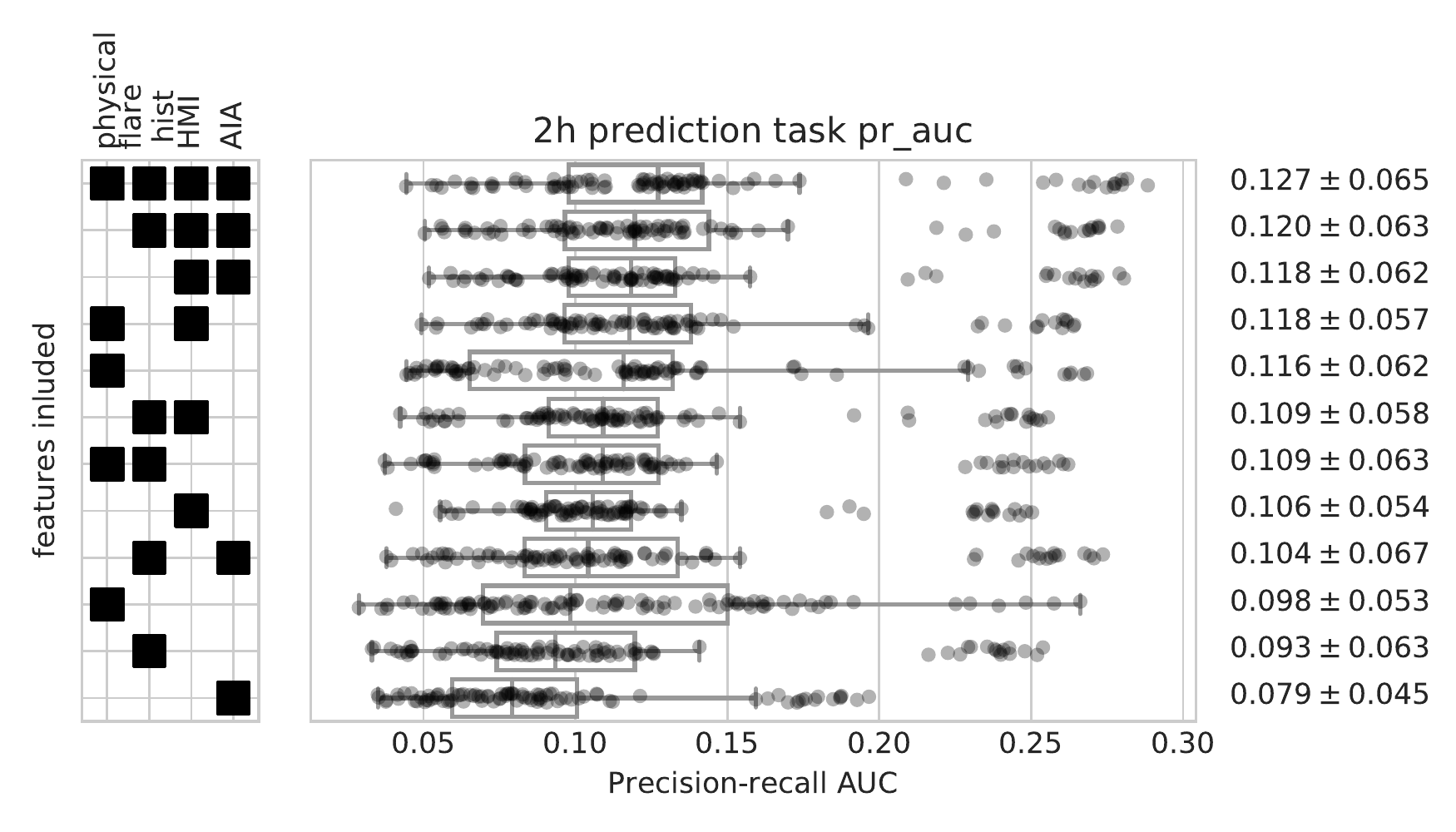} &
\includegraphics[angle=0,width=0.496\textwidth]{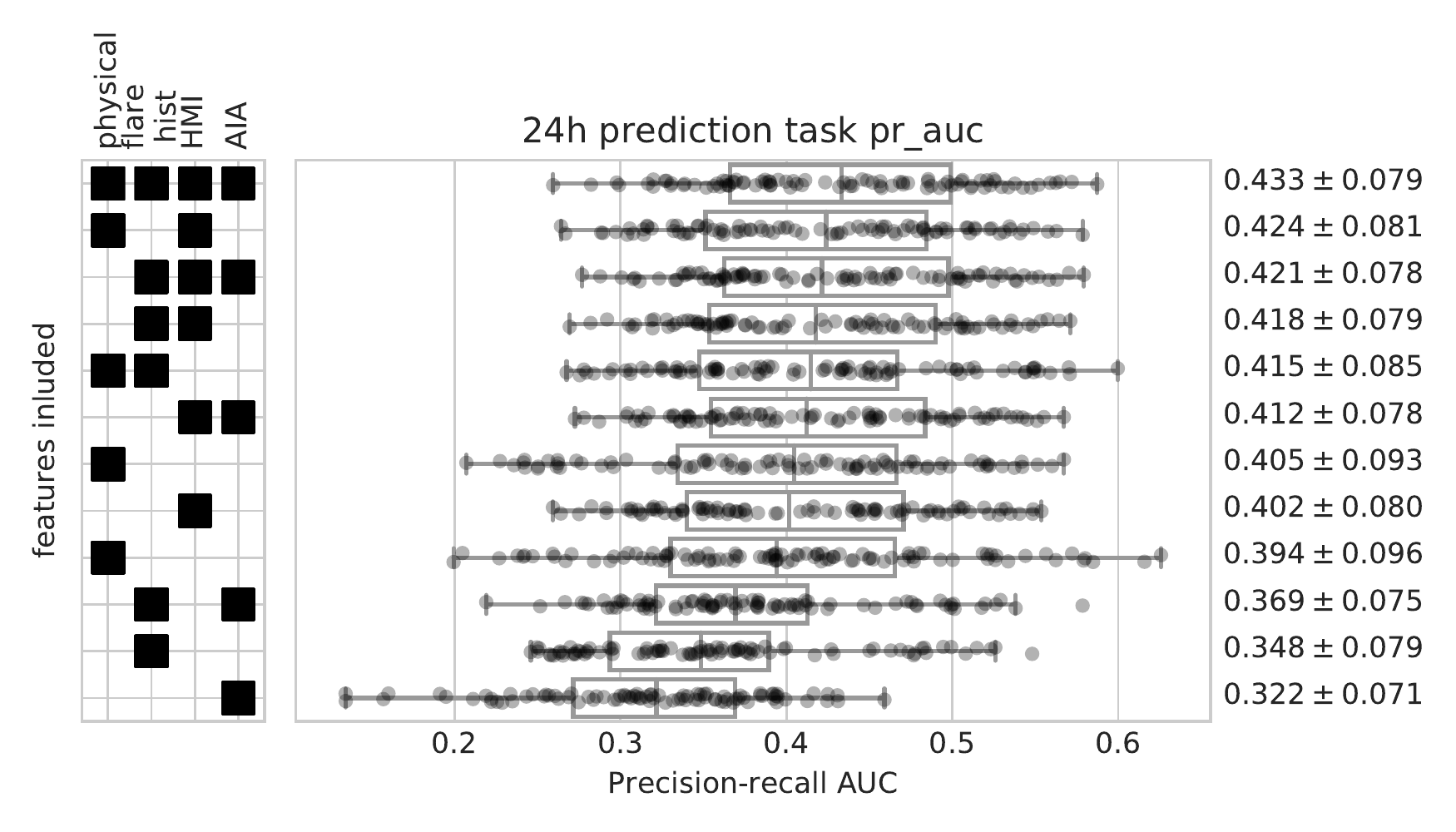}\\
\end{tabular}
\caption{The performance, as measured by the TSS and area under the precision-recall curve, for various feature combinations (sorted in order of decreasing median performance across 90 folds). We hand-engineer the physical and flaring history features (as described in Sections \ref{section:features:physical} and \ref{section:features:flarehist}, respectively) and automatically derive features from the HMI and AIA image data (described in Section \ref{section:features:imagefeatures}). All these features are then used in conjunction with the linear classifier across all tasks and metrics. In general, we find that when optimizing for the TSS, photospheric vector magnetogram data combined with flaring history yields the best performance, and when optimizing for the area under the precision-recall curve, all the data are helpful.}
\label{fig:all_compare_baseline}
\end{figure*}

The tremendous variance in predicted performance, as shown in Figure \ref{fig:all_compare_baseline}, makes it difficult to definitively say if one technique or feature dominated another, but broad comparisons are possible. In general, we find that when optimizing for the TSS, photospheric vector magnetogram data combined with flaring history yields the best performance, and when optimizing for the area under the precision-recall curve, all the data are helpful.

Robustness is vital in assessing if machine learning methods can be useful in operational settings. As such, we focus extensively on cross-validation and use at least 90 folds to obtain our results. In particular, these cross-validation attempts are consistent -- we save the random seed and thus ensure that every classifier is trained on the \textit{same} set of 90 folds. No fold ever has data from the same HARP in both train and test sets. This mimics the real world, in which the goal would be to predict flaring activity for a new, previously unseen active region. Our method is entirely causal -- since we only look at flaring history and instantaneous images (images taken at the current time), there is no possibility for future data or flaring activity to influence our predictions. 

We find, broadly speaking, that our results are only slightly better than the original results presented in \cite{Bobra2015}. While it is common in machine learning literature to celebrate a 1\% improvement on any given metric, the utility of this small of improvement is questionable in practice.

\section{Discussion}
\label{section:conclusion}

We developed an algorithm that automatically generates features in 5.5 TB of SDO image data taken during the time period between May 2010 and May 2014, combines these with other features that describe the flaring history of an active region and physical properties of the photospheric vector magnetic field, and uses a linear classifier on all these features to predict whether an active region will flare within $T$ hours, where $T$ = 2 and 24. Such an approach is conceptually similar to a single-layer Convolutional Neural Network (CNN) with pre-specified filters that is trained using a linear classifier. This is the first attempt to predict solar flares using photospheric vector magnetic field data as well as multiple wavelengths of image data from the chromosphere, transition region, and corona.

Our relatively high scores are similar to, but not greater than, other attempts to predict solar flares (e.g. \citealt{barnes08}, \citealt{mason10}, \citealt{falconer12}, \citealt{ahmed13}, \citealt{boucheron15}). Machine-learning appears to be a robust method for flare prediction, with various different models such as a Support Vector Machine, Lasso linear regression, random forests, and nearest neighbors (e.g. \citealt{song09}, \citealt{yu09}, \citealt{Bobra2015}, \citealt{nishizuka17}) all delivering a higher performance than that reported by the National Oceanic and Atmospheric Administration's Space Weather Prediction Center \citep{crown12}. Given the similar values of algorithm performance across various types of predictive models reported in the literature, we conclude that we can expect a baseline TSS value at least $\sim 0.7$ using SDO data. We also conclude that features automatically detected in HMI image data perform as well as hand-engineered features in predicting solar flares. However, we were surprised to find that adding substantially more data, from multiple diverse data sources, than previous studies did not result in a significantly higher predictive performance. We speculate that there may be many reasons for this.

While our task differs slightly from previous tasks (in predicting a forward-looking window), we still treat the problem as one of binary classification. This may in fact be throwing away tremendous structure in the data. Explicitly predicting the time and magnitude of a future flaring event may ultimately prove more powerful. It may also be that commingling M- and X-class flares, which span multiple orders of magnitude in intensity, into the positive class may be confusing potentially different physical processes.

Rare event prediction is fundamentally challenging, and large class imbalances such as ours remain an active area of research. Additionally, a prediction model is memoryless in that it does not consider previous predictions when making subsequent predictions. Adding an autoregressive component to our predictions may improve accuracy. 

It may be that we have reached the limits of flare prediction accuracy for the prediction windows of 2 and 24 hours. Another avenue for future work is using higher-cadence HMI and AIA data to predict flaring events on extremely short scales such as tens of minutes. Folding in the time variation of the physical features on these scales may help well. The reliability and low latency of the SDO data products would potentially enable the operationalization of this type of prediction.

Astrophysical data sets are full of complexity not present in the type of image data commonly used with CNNs and similar algorithms that automatically detect features in image data through learned filters. For example, the AIA EUV image data used in this study spans nearly five orders of magnitude in dynamic range. The HMI photospheric magnetic field images map a vector, not scalar, field. Inferring the vector magnetic field on the solar surface is no easy task, and, as a result, certain elements of the magnetic field maps contain vastly different signal-to-noise properties than others. As such, the application of CNN-type models must be specifically tailored to take such characteristics of astrophysical data sets into account. 

\acknowledgments
The data used here are courtesy of the GOES team and the NASA/SDO and the HMI and AIA science teams. This work was supported by NASA Grant NAS5-02139 (HMI), and in part by DHS Award HSHQDC-16-3-00083, NSF CISE Expeditions Award CCF-1139158, DOE Award SN10040 DE-SC0012463, and DARPA XData Award FA8750-12-2-0331, and gifts from Amazon Web Services, Google, IBM, SAP, The Thomas and Stacey Siebel Foundation, Apple Inc., Arimo, Blue Goji, Bosch, Cisco, Cray, Cloudera, Ericsson, Facebook, Fujitsu, HP, Huawei, Intel, Microsoft, Mitre, Pivotal, Samsung, Schlumberger, Splunk, State Farm and VMware. BR is generously supported by NSF award CCF-1359814, ONR awards N00014-14-1-0024 and N00014-17-1-2191, the DARPA Fundamental Limits of Learning (Fun LoL) Program, a Sloan Research Fellowship, and a Google Faculty Award.

%  === END ===

\bibliographystyle{aasjournal}
\bibliography{jonas.bib,bobra.bib}
\end{document}